\documentclass{aa}  

\usepackage{graphicx}
\usepackage{txfonts}
\usepackage{booktabs}
\usepackage{placeins}
\usepackage[]{hyperref}
\begin{document} 

   \title{VISTA Magellanic Clouds deep observations of Classical Cepheids:} 
   \subtitle{Precision distances and a collision-driven SMC}

   \author{V. Ripepi \inst{1} 
   \and R. Molinaro \inst{1} 
   \and M. Marconi \inst{1} 
   \and M.-R. L. Cioni \inst{2} 
   \and J. Th. van Loon \inst{3}
   \and R. de Grijs\inst{4,5}
   \and A. Omkumar \inst{2}
   \and J. Storm \inst{2} 
   \and T. Sicignano\inst{1,6,7,8}
   \and J. Oliveira \inst{2}
   \and V. Ivanov \inst{6}
   \and M. A. T. Groenewegen \inst{9}
   \and G. De Somma \inst{1,8}
   \and F. Cusano \inst{10}
   }

   \institute{INAF-Osservatorio Astronomico di Capodimonte, salita Moiariello 16, I-80131 Napoli, Italy 
   \and Leibniz-Institut für Astrophysik Potsdam, An der Sternwarte 16, D-14482 Potsdam, Germany
\and Lennard-Jones Laboratories, Keele University, ST5 5BG, UK
   \and School of Mathematical and Physical Sciences, Macquarie University, Balaclava Road, Sydney, NSW 2109, Australia
\and Astrophysics and Space Technologies Research Centre, Macquarie University, Balaclava Road, Sydney, NSW 2109, Australia
\and 
European Southern Observatory, Karl-Schwarzschild-Strasse 2, 85748 Garching bei München, Germany 
\and 
Scuola Superiore Meridionale, Largo San Marcellino10 I-80138 Napoli, Italy 
\and Istituto Nazionale di Fisica Nucleare, Sez. di Napoli, Monte S. Angelo, Via Cinthia Edificio 6 I-80126 Napoli, Italy
\and 
Koninklijke Sterrenwacht van Belgi\"e, Ringlaan 3, B-1180 Brussels, Belgium
\and
INAF-Osservatorio di Astrofisica e Scienza dello Spazio, Via Gobetti 93/3, I-40129 Bologna, Italy 
}

   \date{Received September 15, 1996; accepted March 16, 1997}


\abstract
  {The central regions of the Magellanic Clouds are crucial for understanding galaxy interactions and calibrating the cosmic distance scale, yet they remain observationally challenging due to extreme stellar crowding.}
  {We utilised deep VISTA observations (VMCDeep) to determine the highest-precision near-infrared Period-Luminosity (PL) and Period-Wesenheit (PW) relations to date for Classical Cepheids ($\delta$ Cep class, DCEPs) in the dense central regions of the Large and Small Magellanic Clouds (LMC and SMC).}
  {We analysed $J$ and $K_s$ time-series data covering the central $1.77~\text{deg}^{2}$ of each galaxy. Utilising approximately 40 epochs per filter and PSF-based photometry, we derived high-quality light curves for 573 and 2090 DCEPs in the LMC and SMC, respectively.}
  {We derived exceptionally precise PL and PW relations, exhibiting the lowest photometric dispersion published so far for the Magellanic Clouds. Fourier parameters calculated from the light curves indicate that pulsational models incorporating a mildly non-canonical mass-luminosity relation and a mixing-length parameter of $\alpha=1.7$ best reproduce the observations. Age derivations confirm a bimodal star formation history in the SMC centre. Crucially, we discovered a striking line-of-sight spatial dichotomy within the young SMC DCEP population ($t < 116$~Myr), which is bifurcated into two distinct structures separated by $\sim 4$~kpc. This provides independent stellar evidence tracing a recent direct collision with the LMC. Finally, by unifying fundamental and first-overtone pulsators, we derived a highly constrained relative distance between the Clouds. Comparing this to detached eclipsing binary distances, our data suggest a moderate metallicity correction for the Leavitt Law with $\gamma \approx -0.2~\text{mag~dex}^{-1}$.}
  {}

               
   \maketitle
%

\section{Introduction}

The Large and Small Magellanic Clouds (LMC and SMC, MCs) represent our nearest pair of interacting dwarf irregular galaxies, currently located at distances of approximately 50--60 kpc \citep[e.g.][]{degrijs2014, degrijs2015}. Due to their proximity and the possibility of studying their stars in great detail, the LMC and SMC are important objects for a variety of astrophysical topics. In particular, the MCs (especially the LMC) serve as primary anchors for calibrating the cosmic distance ladder \citep[e.g.][and references therein]{Freedman2010,Freedman2011,Riess2016}. Among the various tracers, Classical Cepheids ($\delta$ Cep class, DCEPs hereafter) are of paramount importance, due to the tight correlation between their pulsation periods and luminosities or Leavitt Law \citep[][]{Leavitt1912,Freedman2010}, which makes them the most relevant standard candles in the cosmic distance scale and eventually to the estimate of the Hubble constant $H_0$ \citep[e.g.][]{Riess2016,Riess2022a}. 
In addition, DCEPs are excellent tracers of the young stellar population (between 50--300 Myrs) both in the Milky Way (MW) and in the MCs. Indeed, they have been extensively used to trace the spiral arms of the MW \citep[e.g.][and references therein]{Chen2019,GaiaDrimmel2023,Drimmel2025} and the Galactic radial abundance gradients \citep[e.g.][and references therein]{Luck2018,daSilva2023,trentin2024,Trentin2026}. In the MCs, the DCEPs have been employed to calculate the viewing angles and the three-dimensional geometry of both the LMC and SMC \citep[e.g.][and references therein]{2015A&A...573A.135S,Jacy2016,Ripepi2017,Ripepi2022b}.

However, observing individual stars in the central regions of these galaxies is notoriously difficult. The inner parts are prone to extreme crowding and host significant amounts of gas and dust \citep[see][]{Cioni2026}. Large-scale surveys often lose completeness in these high-density regions; for example, the Survey of the Magellanic Clouds Stellar History (SMASH) shows a distinct lack of stars in the LMC centre due to crowding effects \citep[][]{nidever2017, Choi2018}, while the STEP survey \citep[the SMC in Time: Evolution of a Prototype interacting late-type dwarf galaxy][]{Ripepi2014a,Gatto2021} discussed the effect of crowding in the SMC. Even the \textit{Gaia} mission exhibits at least 50\% incompleteness within a 2.5 deg radius of the LMC centre \citep[]{jimenez2023, rathore2025}. Furthermore, optical monitoring by the Optical Gravitational Lensing Experiment (OGLE) has faced challenges where blending effects can mimic line-of-sight extensions \citep[]{Jacy2016, Cusano2021}.

The VISTA survey of the Magellanic Clouds system (VMC) was designed to probe these dusty, dense regions with near-infrared (NIR) imagery at high sensitivity and high spatial resolution \citep[]{Cioni2011}. While the VMC survey has revolutionised our view of the Clouds, the standard survey strategy remained limited in the densest tiles, where sources were still blended and faint variables remained undetected \citep[]{Cioni2026}.

To overcome these hurdles, the VMCDeep programme was initiated to obtain deeper imaging of the inner regions \citep[]{Cioni2026}. By focusing on a single VISTA tile ($\sim$1.77 deg$^2$) per galaxy centre, with significantly better sky quality and an increased number of epochs, this programme provides the most homogeneous NIR spatial coverage of the Clouds' centres to date \citep[e.g][]{Cioni2026}. 

In this paper, we focus on the DCEPs population within these central tiles to determine the first high-precision NIR Period--Luminosity (PL) and Period--Wesenheit\footnote{The Wesenheit magnitudes are reddening-free by construction if the reddening law is known \citep{Madore1982}.} (PW) relations in these crowded regions. The structure of the paper is the following: in Sect. 2, we briefly describe the observations and data analysis; in Sect. 3, we detail the treatment of the light curves to extract the relevant parameters; in Sect. 4, we derive the PL and PW relations for both the LMC and SMC centres; in Sect. 5, we exploit our data in the light of up-to-date pulsational models, and we use our PL and PW relations to constrain their dependence on metallicity; a summary (Sect. 6) closes the paper.

   \begin{figure}
   \centering
   \includegraphics[angle=0,width=8.5cm]{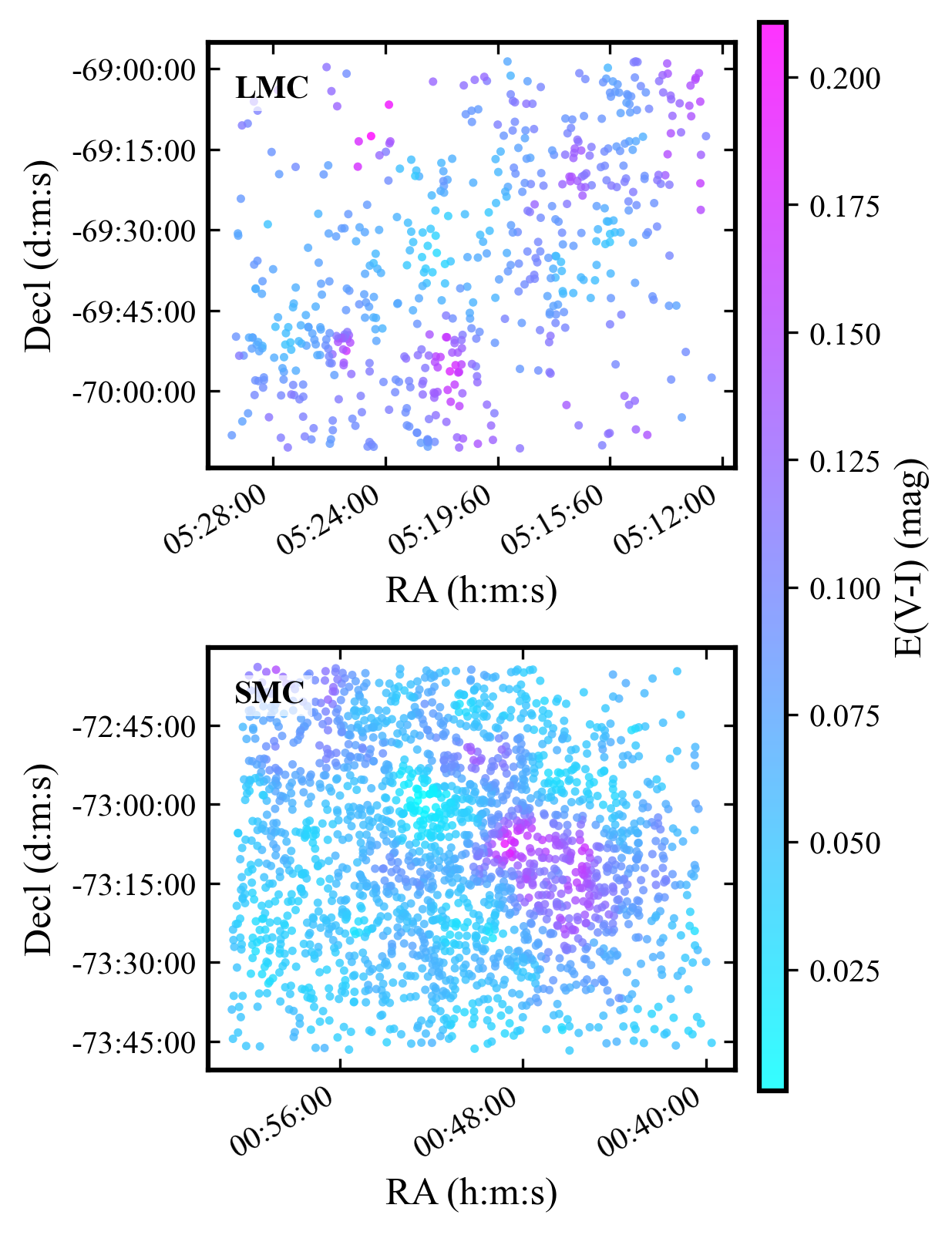}
      \caption{Distribution of DCEPs in the LMC and SMC. The dots are colour-coded according to the estimated reddening $E(V-I)$. For a larger-scale location of the vmcDeep pointings, see \citet{Cioni2026}.}
         \label{fig:radec}
   \end{figure}

\begin{table}[h]
\centering
\caption{Time series for the DCEPs used in this work.}
\label{tab:timeSeries}
\footnotesize\setlength{\tabcolsep}{3pt}
\begin{tabular}{llrrrr}
\hline
  \multicolumn{1}{c}{name} &
  \multicolumn{1}{c}{band} &
  \multicolumn{1}{c}{hjd} &
  \multicolumn{1}{c}{mag} &
  \multicolumn{1}{c}{$\sigma_{\rm mag}$} &
  \multicolumn{1}{c}{tag} \\
    \multicolumn{1}{c}{} &
  \multicolumn{1}{c}{} &
  \multicolumn{1}{c}{(days)} &
  \multicolumn{1}{c}{(mag)} &
  \multicolumn{1}{c}{(mag)} &
  \multicolumn{1}{c}{} \\
\hline
  OGLE-LMC-CEP-1235 & $K_\mathrm{s}$ & 59966.62756 & 16.344 & 0.010 & 0\\
  OGLE-LMC-CEP-1235 & $K_\mathrm{s}$ & 59934.72826 & 16.341 & 0.023 & 0\\
  OGLE-LMC-CEP-1235 & $K_\mathrm{s}$ & 59951.63156 & 16.350 & 0.013 & 0\\
  OGLE-LMC-CEP-1235 & $K_\mathrm{s}$ & 59889.72328 & 16.329 & 0.015 & 0\\
  OGLE-LMC-CEP-1235 & $K_\mathrm{s}$ & 59889.75259 & 16.315 & 0.016 & 0\\
\hline
\end{tabular}
\tablefoot{The meaning of the different columns is: name of the star, band ($J$ or $K_\mathrm{s}$), Heliocentric Julian Day, magnitude and its error, tag indicating data used or not used (outlier) in the fitting of the light curve with 0 and 1, respectively. 
Only the first 5 rows are shown for guidance regarding table form and content. The machine-readable version of the full table will be published at the CDS (Centre de Données astronomiques de Strasbourg, https://cds.u-strasbg.fr/).}
\end{table}

\begin{table}[h]
\centering
\caption{Number of pulsators per mode and galaxy used in this work.}
\label{tab:counts}
\footnotesize\setlength{\tabcolsep}{3pt}
\begin{tabular}{lccccccccc}
\hline
Galaxy & F & 1O & 2O & F1O & 1O2O & Other & Total \\
\hline
LMC & 318 & 213 & 4 & 5 & 29 & 4 & 573 \\
SMC & 1212 & 694 & 35 & 34 & 115 & 0  & 2090 \\
\hline
\end{tabular}
\tablefoot{F, 1O, 2O stand for fundamental, first and second overtone pulsators, respectively. F1O and 1O2O are fundamental--first overtone and first--second overtone mixed modes, respectively. Other indicates rarer mixed mode DCEPs. }
\end{table}

   \begin{figure*}
   \centering
   \includegraphics[angle=0,width=15cm]{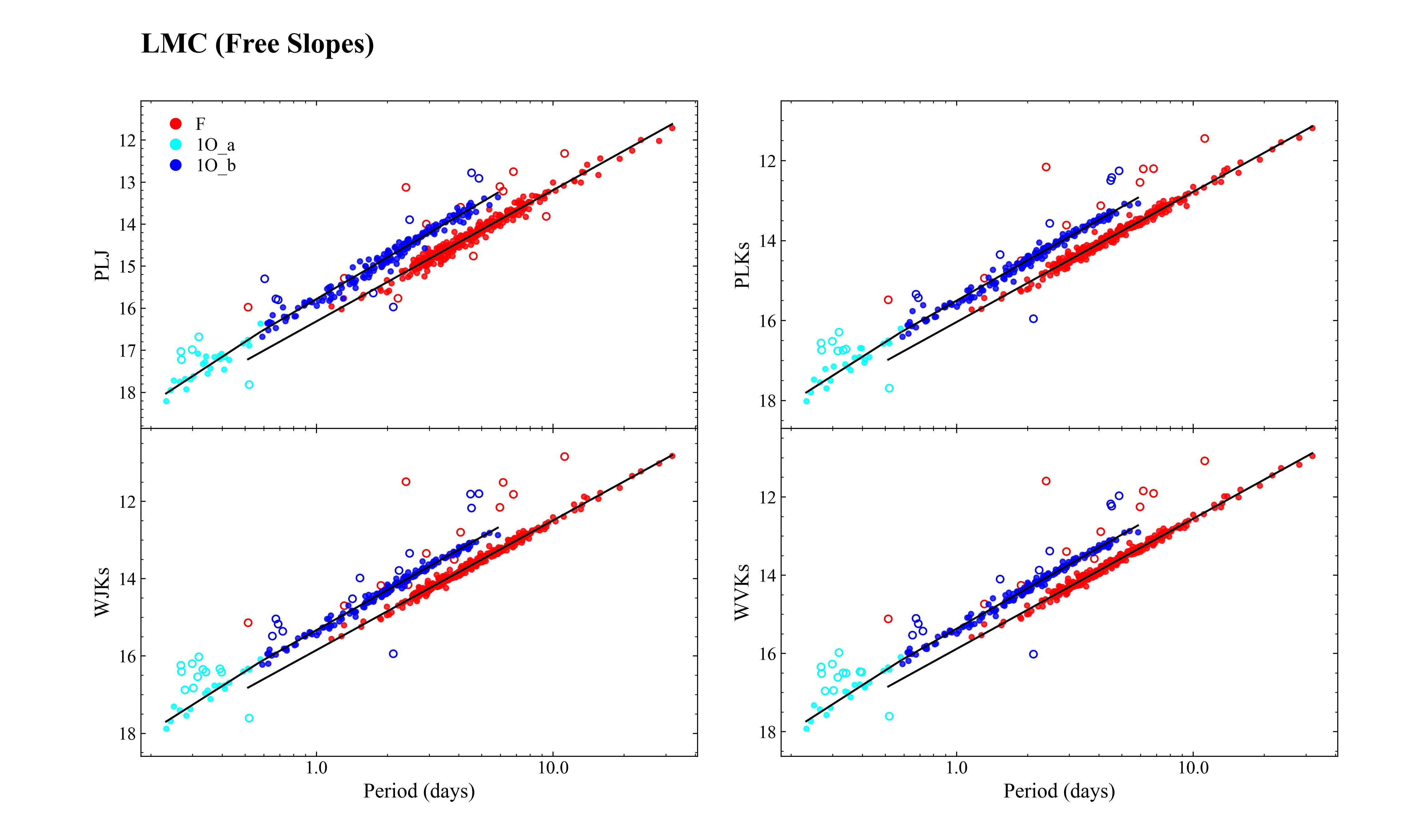}
      \caption{PL and PW relations in $J$ and $K_\mathrm{s}$ for the LMC. Red, blue, cyan and grey filled circles represent F-mode, 1O-mode with period > 0.58d, 1O-mode with period $\leq$ 0.58d, 2O-mode, respectively. The open circles show the outlying DCEPs which are rejected by $\sigma$-clipping during the fitting procedure (see Sect.~\ref{sect:pl}).}
         \label{fig:PL_LMC}
   \end{figure*}

   \begin{figure*}
   \centering
   \includegraphics[angle=0,width=15cm]{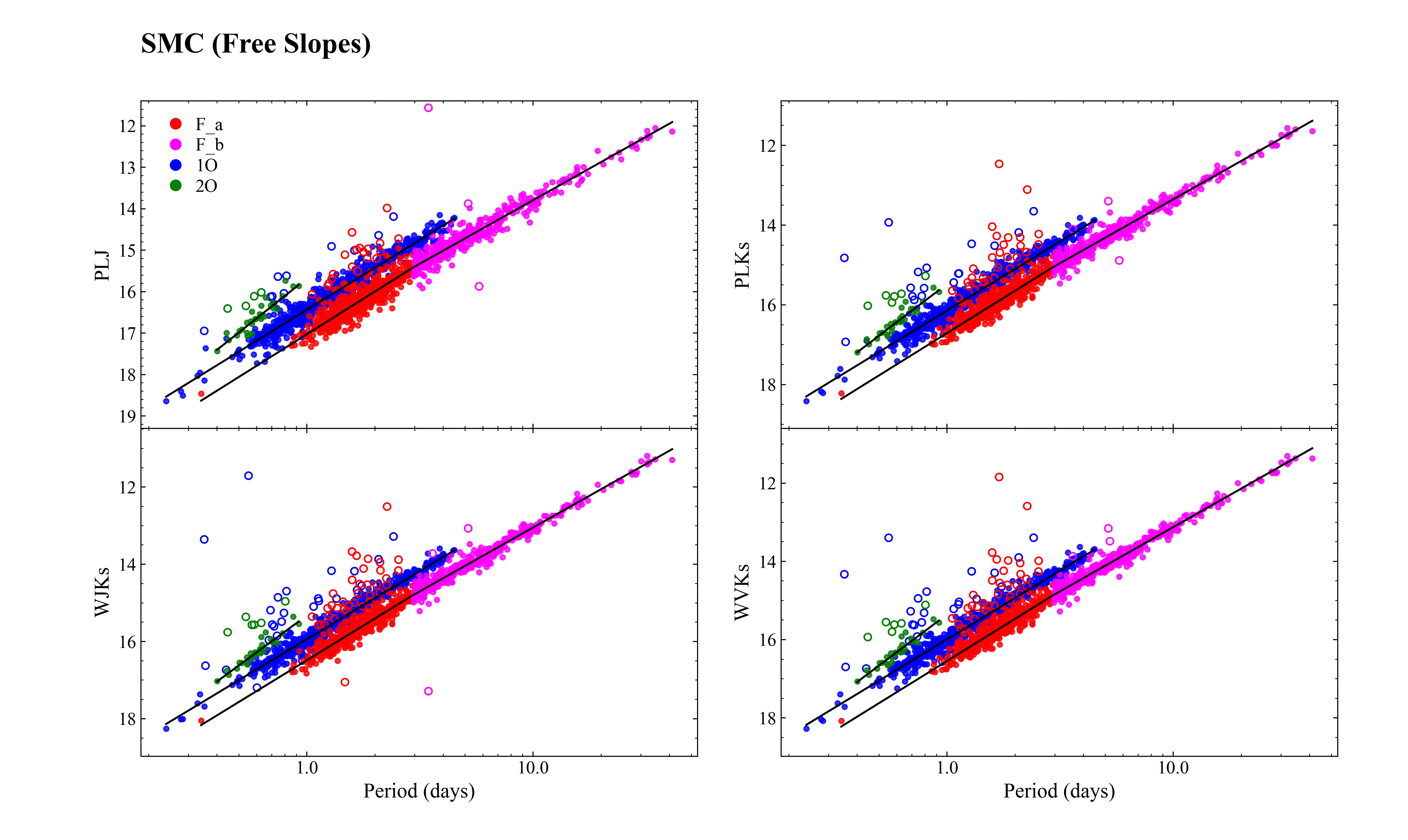}
      \caption{As in Fig.~\ref{fig:PL_LMC} but for the SMC.}
         \label{fig:PL_SMC}
   \end{figure*}

\begin{table*}
\caption{Results from the analysis of the light curves for the DCEPs analysed in this paper.}
\label{table:main_results}
\centering
\footnotesize\setlength{\tabcolsep}{3pt}
\begin{tabular}{l c c c c c c c c c c c c c}
\hline\hline
Name & RA & Decl & Gal. & Mode & $P$ & $N_{\mathrm{pt},J}$ & $N_{\mathrm{out},J}$ & $J$ & $\sigma_J$ & $A_J$ & $\sigma_{A_J}$ & $\sigma_{\mathrm{obs},J}$ & $\mathrm{rms}_J$ \\
 & (deg) & (deg) &  &  & (days) &  &  & (mag) & (mag) & (mag) & (mag) & (mag) & (mag) \\
(1) & (2) & (3) & (4) & (5) & (6) & (7) & (8) & (9) & (10) & (11) & (12) & (13) & (14) \\
\hline
OGLE-SMC-CEP-1918 & 13.14829 & $-$72.98914 & SMC & F & 5.253997 & 64 & 8 & 13.987 & 0.002 & 0.248 & 0.010 & 0.076 & 0.021 \\
OGLE-SMC-CEP-1877 & 13.09133 & $-$72.97006 & SMC & F & 1.438308 & 64 & 1 & 16.273 & 0.009 & 0.578 & 0.028 & 0.190 & 0.090 \\
OGLE-SMC-CEP-1913 & 13.14383 & $-$72.97017 & SMC & F & 15.509895 & 64 & 3 & 13.158 & 0.001 & 0.349 & 0.005 & 0.113 & 0.011 \\
OGLE-SMC-CEP-1772 & 12.98163 & $-$72.96483 & SMC & F & 2.581575 & 64 & 4 & 15.482 & 0.001 & 0.486 & 0.004 & 0.136 & 0.008 \\
OGLE-SMC-CEP-1906 & 13.13312 & $-$73.02022 & SMC & 1 & 2.906754 & 64 & 12 & 14.921 & 0.001 & 0.153 & 0.002 & 0.054 & 0.006 \\
\hline
\end{tabular}

\vspace{0.4cm}

\begin{tabular}{c c c c c c c c c c c c c c c c}
\hline\hline
$\mathrm{qf}_J$ & $\mathrm{UI}_J$ & $N_{\mathrm{pt},K_\mathrm{s}}$ & $N_{\mathrm{out},K_\mathrm{s}}$ & $K_\mathrm{s}$ & $\sigma_{K_\mathrm{s}}$ & $A_{K_\mathrm{s}}$ & $\sigma_{A_{K_\mathrm{s}}}$ & $\sigma_{\mathrm{obs},K_\mathrm{s}}$ & $\mathrm{rms}_{K_\mathrm{s}}$ & $\mathrm{qf}_{K_\mathrm{s}}$ & $\mathrm{UI}_{K_\mathrm{s}}$ & $V$ & $E(V-I)$ & Age & $\sigma$Age\\
 &  &  &  & (mag) & (mag) & (mag) & (mag) & (mag) & (mag) &  &  & (mag) & (mag) & (Myr) & (Myr)\\
(15) & (16) & (17) & (18) & (19) & (20) & (21) & (22) & (23) & (24) & (25) & (26) & (27) & (28) & (29) & (30)\\
\hline
3.64 & 0.979 & 50 & 2 & 13.689 & 0.003 & 0.184 & 0.012 & 0.059 & 0.021 & 2.88 & 0.975 & 15.228 & 0.001 & 110& 10\\
2.1 & 0.983 & 50 & 3 & 15.924 & 0.007 & 0.445 & 0.027 & 0.135 & 0.068 & 2 & 0.976 & 17.395 & 0.006  & 211& 19\\
10.72 & 0.969 & 50 & 2 & 12.675 & 0.001 & 0.332 & 0.004 & 0.105 & 0.013 & 8.18 & 0.966 & 14.723 & 0.007  & 120& 37\\
16.68 & 0.975 & 50 & 2 & 15.142 & 0.001 & 0.486 & 0.004 & 0.091 & 0.008 & 11.13 & 0.985 & 16.509 & 0.010  & 149& 45\\
8.44 & 0.958 & 50 & 1 & 14.575 & 0.001 & 0.090 & 0.002 & 0.033 & 0.008 & 4.09 & 0.978 & 16.035 & 0.011  & 81& 22\\
\hline
\end{tabular}

\tablefoot{ The meaning of the columns is: 
(1) Target identification from the OGLE survey; 
(2) \& (3) Right Ascension and Declination (J2000); 
(4) Host galaxy designation (LMC/SMC); 
(5) Pulsation mode identifier: F=Fundamental, 1=First overtone, 2=second overtone, any combination of these modes indicates multimode pulsators; 
(6) Pulsation period; 
(7) \& (8) Number of total light curve points and rejected outliers in the $J$ band; 
(9) Mean intensity-magnitude in the $J$ band; 
(10) Mean magnitude uncertainty in $J$; 
(11) \& (12) Peak-to-peak amplitude of the fit model in the $J$ band and its error; 
(13) Standard deviation of the observational data points in $J$; 
(14) Root-mean-square residual of the light curve fit in $J$; 
(15) Light curve quality factor in $J$; 
(16) Phase uniformity index in $J$; 
(17)-(26) Same as (7)—(16) but for the $K_\mathrm{s}$ band;
(27) Average optical $V$-band magnitude (where a value of 99.999 indicates missing or unavailable data); 
(28) Adopted $E(V-I)$ interstellar reddening value;
(29) Age in Myr;
(30) Error on Age in Myr.
Only the first 5 rows are shown for guidance regarding table form and content. The machine-readable version of the full table will be published at the CDS (Centre de Données astronomiques de Strasbourg, https://cds.u-strasbg.fr/).
}
\end{table*}

\begin{table*}
\centering
\footnotesize\setlength{\tabcolsep}{3pt}
\caption{Parameters of the PL and PW relations for LMC and SMC Cepheids. }
\label{tab:pl_pw_results}
\begin{tabular}{llc|ccccc|ccccc}
\hline
 {Relation} &  {Galaxy} &  {Sample} & ${a}$ &  ${b}$ &  {rms} & ${N_{\rm pts}}$ & ${N_{\rm out}}$ &  ${a}$ &  ${b}$ &  {rms} & ${N_{\rm pts}}$ & ${N_{\rm out}}$\\
   &          &         &            (mag)   & (mag dex$^{-1}$)      &   (mag)  &  &  &  (mag)   & (mag dex$^{-1}$)      &   (mag)  &  & \\  
\hline
\multicolumn{3}{c|}{} & \multicolumn{5}{c}{Free Slopes} & \multicolumn{5}{|c}{Fixed Slopes}\\
\hline
PLJ & LMC & F & $16.307 \pm 0.052$ & $-3.109 \pm 0.096$ & 0.111 & 314 & 14      &   $16.294 \pm 0.006$ & $-3.084$ & 0.110 & 314 & 12 \\       
PLKs & LMC & F & $16.035 \pm 0.038$ & $-3.249 \pm 0.071$ & 0.084 & 314 & 11  	&   $16.024 \pm 0.005$ & $-3.230$ & 0.085 & 314 & 10 \\      
WJKs & LMC & F & $15.845 \pm 0.033$ & $-3.348 \pm 0.061$ & 0.072 & 314 & 11  	&   $15.826 \pm 0.004$ & $-3.314$ & 0.072 & 314 & 12 \\      
WVKs & LMC & F & $15.892 \pm 0.034$ & $-3.336 \pm 0.063$ & 0.072 & 314 & 12  	&   $15.874 \pm 0.004$ & $-3.303$ & 0.073 & 314 & 11 \\      
PLJ & LMC & 1O\_a & $15.698 \pm 0.019$ & $-3.873 \pm 0.069$ & 0.113 & 30 & 6  	   &   $15.684 \pm 0.008$ & $-3.773$ & 0.119 & 30 & 6 \\     
PLJ & LMC & 1O\_b & $15.795 \pm 0.019$ & $-3.333 \pm 0.089$ & 0.113 & 195 & 8  	   &   $15.791 \pm 0.008$ & $-3.319$ & 0.119 & 195 & 8 \\    
PLKs & LMC & 1O\_a & $15.422 \pm 0.016$ & $-3.967 \pm 0.057$ & 0.094 & 30 & 8  	    &   $15.409 \pm 0.007$ & $-3.867$ & 0.098 & 30 & 8 \\    
PLKs & LMC & 1O\_b & $15.523 \pm 0.016$ & $-3.415 \pm 0.073$ & 0.094 & 195 & 10     &   $15.519 \pm 0.007$ & $-3.402$ & 0.098 & 195 & 10 \\   
PWJKs & LMC & 1O\_a & $15.228 \pm 0.012$ & $-4.043 \pm 0.043$ & 0.072 & 30 & 13      &   $15.222 \pm 0.005$ & $-3.943$ & 0.075 & 30 & 13 \\   
PWJKs & LMC & 1O\_b & $15.340 \pm 0.012$ & $-3.468 \pm 0.056$ & 0.072 & 195 & 12     &   $15.337 \pm 0.005$ & $-3.457$ & 0.075 & 195 & 12 \\   
PWVKs & LMC & 1O\_a & $15.284 \pm 0.013$ & $-3.966 \pm 0.047$ & 0.078 & 30 & 12      &   $15.272 \pm 0.006$ & $-3.866$ & 0.081 & 30 & 12 \\   
PWVKs & LMC & 1O\_b & $15.375 \pm 0.013$ & $-3.454 \pm 0.061$ & 0.078 & 195 & 10     &   $15.372 \pm 0.006$ & $-3.442$ & 0.081 & 195 & 10 \\   
PLJ & SMC & 1O & $16.444 \pm 0.015$ & $-3.384 \pm 0.041$ & 0.178 & 783 & 10  	 &   $16.443 \pm 0.006$ & $-3.377$ & 0.178 & 783 & 10 \\     
PLKs & SMC & 1O & $16.152 \pm 0.012$ & $-3.493 \pm 0.034$ & 0.148 & 783 & 17  	 &   $16.152 \pm 0.005$ & $-3.489$ & 0.148 & 783 & 17 \\     
PWJKs & SMC & 1O & $15.952 \pm 0.011$ & $-3.576 \pm 0.030$ & 0.129 & 783 & 26  	 &   $15.951 \pm 0.004$ & $-3.572$ & 0.129 & 783 & 26 \\     
PWVKs & SMC & 1O & $16.004 \pm 0.011$ & $-3.548 \pm 0.031$ & 0.134 & 783 & 23  	 &   $16.003 \pm 0.004$ & $-3.544$ & 0.134 & 783 & 23 \\     
PLJ & SMC & F\_a & $17.026 \pm 0.012$ & $-3.389 \pm 0.024$ & 0.174 & 830 & 18  	  &   $17.027 \pm 0.005$ & $-3.406$ & 0.175 & 830 & 19 \\    
PLJ & SMC & F\_b & $16.868 \pm 0.012$ & $-3.107 \pm 0.048$ & 0.174 & 368 & 3  	  &   $16.870 \pm 0.005$ & $-3.070$ & 0.175 & 368 & 3 \\     
PLKs & SMC & F\_a & $16.711 \pm 0.010$ & $-3.504 \pm 0.020$ & 0.143 & 830 & 28     &   $16.712 \pm 0.004$ & $-3.513$ & 0.143 & 830 & 29 \\   
PLKs & SMC & F\_b & $16.575 \pm 0.010$ & $-3.220 \pm 0.040$ & 0.143 & 368 & 4  	   &   $16.576 \pm 0.004$ & $-3.224$ & 0.143 & 368 & 4 \\    
PWJKs & SMC & F\_a & $16.495 \pm 0.009$ & $-3.585 \pm 0.018$ & 0.128 & 830 & 37     &   $16.496 \pm 0.003$ & $-3.597$ & 0.128 & 830 & 37 \\   
PWJKs & SMC & F\_b & $16.372 \pm 0.009$ & $-3.344 \pm 0.035$ & 0.128 & 368 & 4  	   &   $16.373 \pm 0.003$ & $-3.334$ & 0.128 & 368 & 4 \\    
PWVKs & SMC & F\_a & $16.548 \pm 0.009$ & $-3.557 \pm 0.018$ & 0.128 & 830 & 35     &   $16.549 \pm 0.003$ & $-3.567$ & 0.128 & 830 & 35 \\   
PWVKs & SMC & F\_b & $16.418 \pm 0.009$ & $-3.300 \pm 0.035$ & 0.128 & 368 & 4  	   &   $16.419 \pm 0.003$ & $-3.291$ & 0.128 & 368 & 4 \\    
 \hline
\end{tabular}
\tablefoot{PL and PW followed by a band or colour names stands for Period--Luminosity or Period--Wesenheit in the specific bands/colours. The fitted relation has the form $M_{\lambda} = a + b \times (\log P - \log P_0)$, where $M_{\lambda}$ is the generic apparent magnitude or Wesenheit quantity, $P$ is the period, and $P_0$ represents the pivoting period adopted to reduce the covariance between $a$ and $b$. These values are the following: LMC: F= 5.0 days; 1O\_a=0.4 days; 1O\_b=2.0 days. 
        SMC: F\_a=10.0 days; F\_b=2.0 days; 1O=2.0 days, 2O=0.6 days. The different pivoting periods for the F-mode pulsators in LMC and SMC are due to the much reduced span in periods in the former galaxy. 
        It is important to note that the zero points shown in the table ($a$ values) have been corrected to remove the pivot period, so they can be used as they are. The left and right parts of the table show results where the slope was left as a free parameter or fixed to literature values, respectively (see Sect.~\ref{sect:pl}).}
\end{table*}

\begin{table*}
\centering
\footnotesize\setlength{\tabcolsep}{3pt}
\caption{Relative distance moduli between the centres of SMC and LMC.}
\label{tab:relative_distances}
\begin{tabular}{llcccccc}
\toprule
 & & \multicolumn{2}{c}{${\gamma = 0.0}$} & \multicolumn{2}{c}{${\gamma = -0.2}$} & \multicolumn{2}{c}{${\gamma = -0.4}$} \\
\cmidrule(lr){3-4} \cmidrule(lr){5-6} \cmidrule(lr){7-8}
 {Paradigm} &  {Relation} &  {Mode} & ${\Delta\mu}$ (mag) &  {Mode} & ${\Delta\mu}$ (mag) &  {Mode} & ${\Delta\mu}$ (mag) \\
\midrule
Free Slopes  & PLJ  & F & $0.573 \pm 0.034$ & F & $0.505 \pm 0.034$ & F & $0.437 \pm 0.035$ \\
             &      & 1O & $0.656 \pm 0.085$ & 1O & $0.588 \pm 0.085$ & 1O & $0.520 \pm 0.086$ \\
             &      & Comb. & $0.584 \pm 0.031$ & Comb. & $0.516 \pm 0.032$ & Comb. & $0.448 \pm 0.033$ \\
             & PLKs & F & $0.546 \pm 0.028$ & F & $0.478 \pm 0.029$ & F & $0.410 \pm 0.030$ \\
             &      & 1O & $0.641 \pm 0.073$ & 1O & $0.573 \pm 0.073$ & 1O & $0.505 \pm 0.074$ \\
             &      & Comb. & $0.558 \pm 0.026$ & Comb. & $0.490 \pm 0.027$ & Comb. & $0.422 \pm 0.028$ \\
             & WJKs & F & $0.528 \pm 0.025$ & F & $0.460 \pm 0.025$ & F & $0.392 \pm 0.026$ \\
             &      & 1O & $0.610 \pm 0.059$ & 1O & $0.542 \pm 0.060$ & 1O & $0.474 \pm 0.060$ \\
             &      & Comb. & $0.540 \pm 0.023$ & Comb. & $0.472 \pm 0.023$ & Comb. & $0.404 \pm 0.024$ \\
             & WVKs & F & $0.544 \pm 0.026$ & F & $0.476 \pm 0.026$ & F & $0.408 \pm 0.027$ \\
             &      & 1O & $0.625 \pm 0.060$ & 1O & $0.557 \pm 0.060$ & 1O & $0.489 \pm 0.061$ \\
             &      & Comb. & $0.557 \pm 0.024$ & Comb. & $0.489 \pm 0.024$ & Comb. & $0.421 \pm 0.025$ \\
\midrule
Fixed Slopes & PLJ  & F & $0.583 \pm 0.008$ & F & $0.515 \pm 0.009$ & F & $0.447 \pm 0.012$ \\
             &      & 1O & $0.652 \pm 0.010$ & 1O & $0.584 \pm 0.011$ & 1O & $0.516 \pm 0.013$ \\
             &      & Comb. & $0.610 \pm 0.006$ & Comb. & $0.542 \pm 0.007$ & Comb. & $0.474 \pm 0.009$ \\
             & PLKs & F & $0.555 \pm 0.006$ & F & $0.487 \pm 0.007$ & F & $0.419 \pm 0.011$ \\
             &      & 1O & $0.633 \pm 0.009$ & 1O & $0.565 \pm 0.010$ & 1O & $0.497 \pm 0.013$ \\
             &      & Comb. & $0.579 \pm 0.005$ & Comb. & $0.511 \pm 0.006$ & Comb. & $0.443 \pm 0.009$ \\
             & WJKs & F & $0.537 \pm 0.005$ & F & $0.469 \pm 0.007$ & F & $0.401 \pm 0.010$ \\
             &      & 1O & $0.614 \pm 0.006$ & 1O & $0.546 \pm 0.008$ & 1O & $0.478 \pm 0.011$ \\
             &      & Comb. & $0.568 \pm 0.004$ & Comb. & $0.500 \pm 0.005$ & Comb. & $0.432 \pm 0.007$ \\
             & WVKs & F & $0.551 \pm 0.005$ & F & $0.483 \pm 0.007$ & F & $0.415 \pm 0.010$ \\
             &      & 1O & $0.631 \pm 0.007$ & 1O & $0.563 \pm 0.008$ & 1O & $0.495 \pm 0.011$ \\
             &      & Comb. & $0.578 \pm 0.004$ & Comb. & $0.510 \pm 0.005$ & Comb. & $0.442 \pm 0.008$ \\
\bottomrule
\end{tabular}
\tablefoot{$\Delta\mu = \mu_{\text{SMC}} - \mu_{\text{LMC}}$. It is evaluated at $\log P_{\text{ref}} = 0.5$ for F mode, $\log P_{\text{ref}} = 0.0$ for 1O mode, and for the fundamentalised combined sample. The geometric DEB distance is $0.500 \pm 0.040$ mag.}
\end{table*}

\section{Observations and data analysis}


The observations analysed in this work were obtained as part of ESO programmes 105.206M, 106.201G, 108.221P, and 109.23G7 between December 2020 and January 2023. We utilised the VIRCAM camera on the 4m VISTA telescope, adopting the $J$ and $K_\mathrm{s}$ filters. The observations focused on two geodesic rectangles centred at $\alpha=05:20:54.47, \delta=-69:34:43.32$ (J2000, LMC) and $\alpha=00:50:16.06, \delta=-73:10:40.92$ (J2000, SMC). All the details about the observational strategy are provided in \citet{Cioni2026}. Here we only remind that the exposure times (360 s and 900 s, in $J$ and $K_\mathrm{s}$, respectively) were designed to detect faint pulsating variable stars (e.g. RR Lyrae and $\delta$ Sct classes) in single-epoch images, whereas the choice of filters reflects the best calibrated VISTA filters, namely $J$ and $K_\mathrm{s}$, which are also those where pulsating stars obey tight PL relations.
The survey strategy for VMCDeep was optimised, among other things, for variability studies by requesting observations spread over different nights, with a seeing of $\sim$0.7 arcsec at 500 nm and approximately 40 epochs for each filter. 

Regarding raw data processing and photometry, we refer the reader to \citet{Cioni2026}. In this study, we adopted the Point Spread Function (PSF) photometry per half-tile, where one half-tile corresponds to the mosaic of three VISTA pawprint images. These three images are designed to fill the gaps among the 16 detectors of the VISTA field-of-view in the vertical direction; a similar set of images shifted also in the horizontal direction results in a homogeneous coverage of the tile. For details about the exposure time and dither pattern that produced a given pawprint image, in a given filter, we refer the reader to \citet{Cioni2026}. The VMCDeep photometry is calibrated in the VISTA photometric system, which is tied to 2MASS \citep[Two-micron All Sky Survey; ][]{Skrutskie2006}, as discussed in detail in \citet[][]{2018MNRAS.474.5459G}.   

All the time series published in this work have been compacted in a unique table for ease of use. The format of the data is exemplified in Table~\ref{tab:timeSeries} which is available electronically at CDS (Centre de Données astronomiques de Strasbourg, https://cds.u-strasbg.fr/).

\subsection{The DCEP Sample and Extinction Correction}
\label{sect:reddening}
The DCEPs analysed in this work were selected by cross-matching our NIR catalogues with the list of DCEPs published by the OGLE survey \citep[][]{Udalski2018,Sos2015,Sos2017}. Specifically, we adopted a 1-arcsec tolerance in the cross-match, yielding a total of 573 and 2090 usable DCEPs in the LMC and SMC, respectively.
The number of targets divided according to the different modes of pulsation is shown in Table~\ref{tab:counts}. From the OGLE catalogue, we adopted the period of pulsation and epochs of maximum light, as well as the $V$ magnitudes. 

For most DCEPs, we have more than the nominal 40 epochs, owing to our choice to also use numerous observations with seeing outside the canonical requirements, but still with a good-enough quality. For example, an image with a seeing of 0.85 arcsec would be considered out of constraints during the creation of the deep stacks \citep[see][]{Cioni2026}, but it provides potentially usable phase points for the light curves of DCEPs, especially the brightest ones (see below). Also, a few tens of objects have almost double the number of epochs, being observed in both half tiles, as described in \citet{Cioni2026}.

One of the aims of this work is the derivation of the PL relations in $J$ and $K_\mathrm{s}$ bands, which requires an evaluation of the reddening for each DCEP. To this end, we adopted the high-resolution (1.7 arcmin $\times$ 1.7 arcmin) empirically calibrated reddening maps of \citet{Skowron2021}, based on the red clump (RC) stars.  We choose these maps for their reliable empirical calibration, relatively high spatial resolution, homogeneity and ease of use. Even if RC stars are known to belong to a different stellar population/environment with respect to DCEPs, these maps agree well with a variety of previous works, obtained using a broad range of different methods \citep[including those based on VMC data by ][only available in the SMC]{Tatton2021}. 
Finally, we interpolated these maps at the DCEPs RA and Dec positions, obtaining an estimate of the $E(V-I)$ value. 

The distribution in the sky of the DCEPs analysed in the LMC and SMC is shown in Fig. ~\ref{fig:radec}. The same figure shows the significantly varying reddening across the two central fields, especially in the SMC, where $E(V-I)$ can reach up to 0.2 mag. 
To correct the NIR magnitudes for the reddening, to be consistent with our previous works, \citep[e.g.][]{Ripepi2016,Ripepi2017,Ripepi2022b}, we used the coefficients from \citet{Cardelli1989}, resulting in:
$J_0=J-0.737\,E(V-I)$ and $K_{s,0}=K_\mathrm{s}-0.30\,E(V-I)$. 
We are aware that more modern reddening laws are available in the literature \citep[see e.g.][]{Fitzpatrick1999,Gordon2023,Wang2023_reddening}, and that different values of $R_V$ are known to hold for LMC and SMC \citep[][]{Gordon2003}, but the largest differences compared with \citet{Cardelli1989} and with $R_V=3.1$ used here, happen at short wavelengths, i.e. in the blue and ultraviolet part of the spectrum. In the NIR regime, on which we based our work, the impact of differences between the different reddening laws in the MCs is negligible \citep[see e.g. sect. 4.1 in][]{Wang2023_reddening}.

   \begin{figure*}
   \centering
   \includegraphics[angle=0,width=14cm]{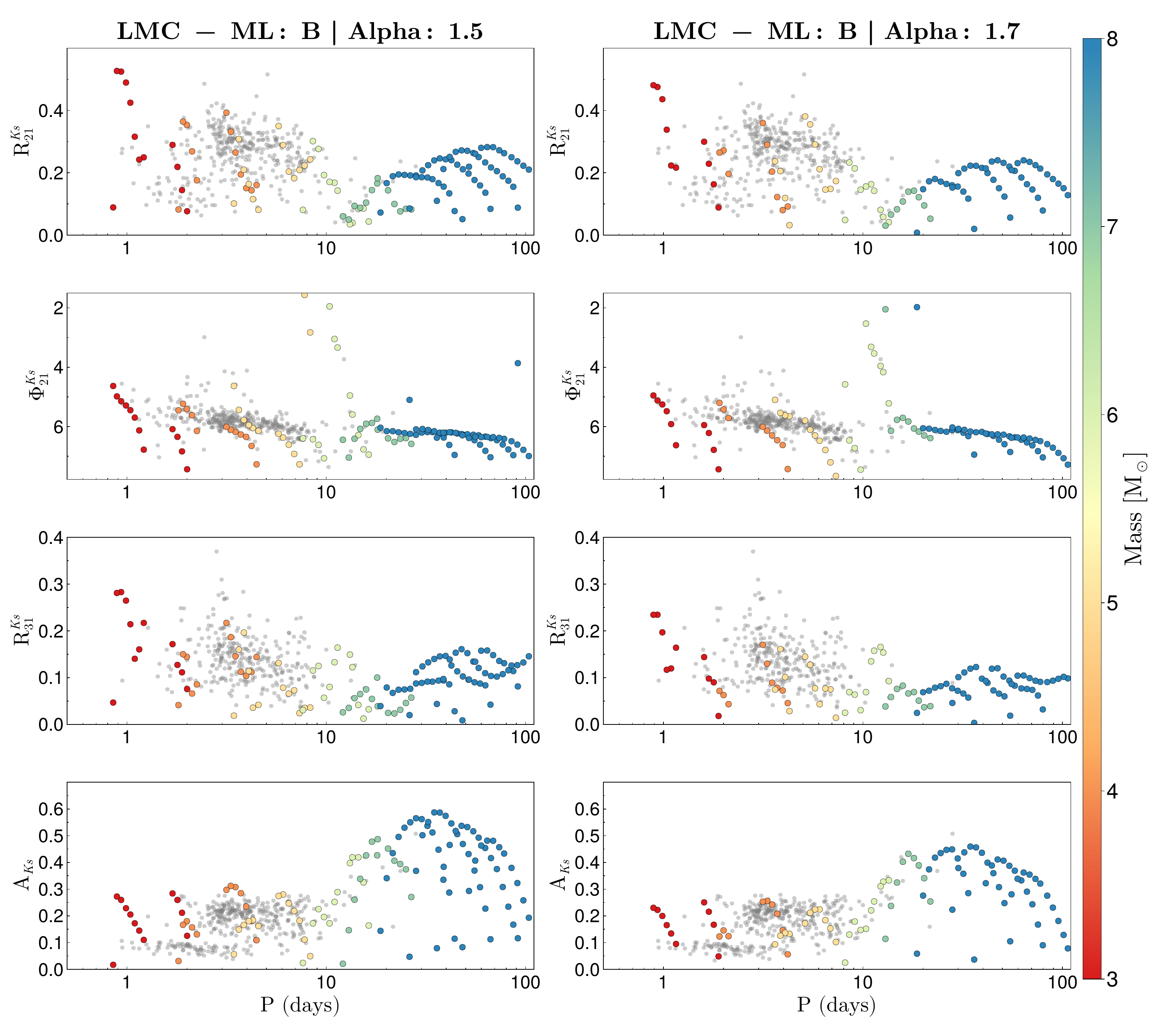}
      \caption{From top to bottom, the figure shows in grey the observed Fourier parameters R$_{21}$, $\Phi_{21}$, R$_{31}$ and the peak-to-peak amplitude (in mag) in the $K_\mathrm{s}$ band for the LMC DCEPs. For comparison, the pulsation models by \citet{Desomma2022} are displayed as filled circles coloured according to their masses, spanning the range 3--8 $M_{\odot}$. The models plotted here ("Set B") have been calculated with a non-canonical ML relation (i.e. with mild overshooting), for two different values of the mixing length parameter $\alpha=$1.5 (left panels) and $\alpha=$1.7 (right panels).}
         \label{fig:fou_lmc}
   \end{figure*}

   \begin{figure*}
   \centering
    \includegraphics[angle=0,width=14cm]{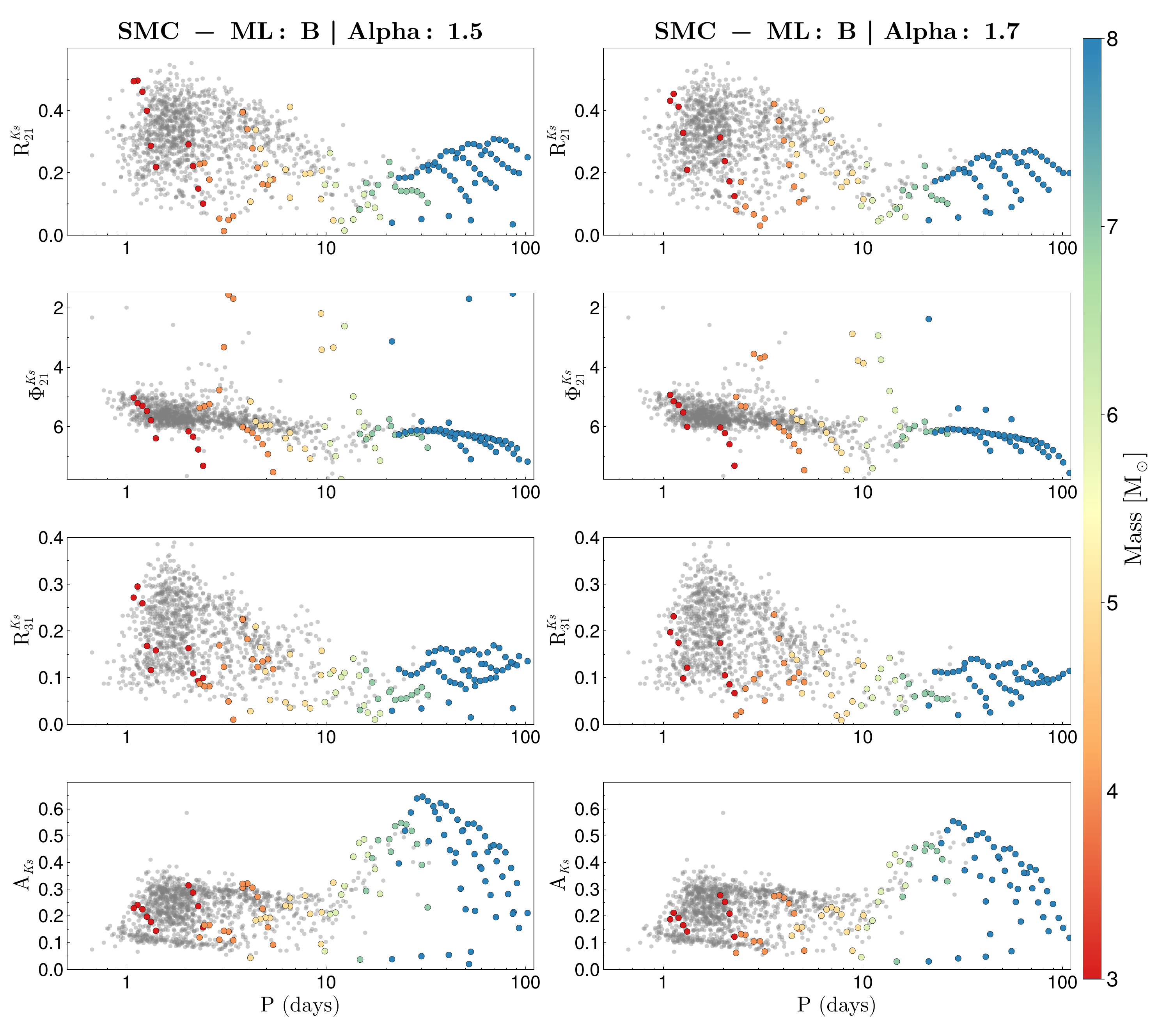} 
      \caption{As in Fig.~\ref{fig:fou_lmc}, but for the SMC.}
         \label{fig:fou_smc}
   \end{figure*}

   \begin{figure}
   \centering
    \includegraphics[angle=0,width=9cm]{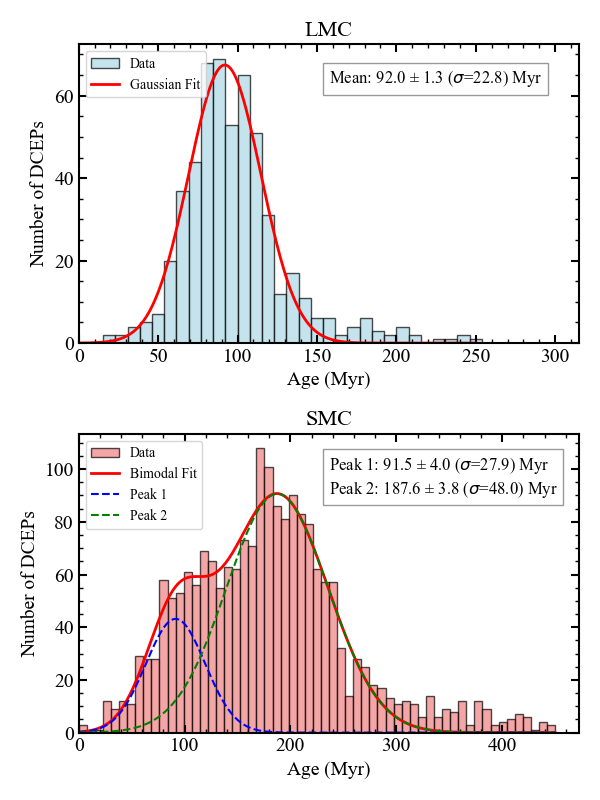} 
      \caption{Age distributions for the target DCEPs in the LMC (top) and SMC (bottom). The solid lines are the Gaussian fits to the distributions (bimodal in the case of SMC). Mean and dispersion of each Gaussian are labelled in the figure.}
         \label{fig:ageHistograms}
   \end{figure}

   \begin{figure}
   \centering
    \includegraphics[angle=0,width=9cm]{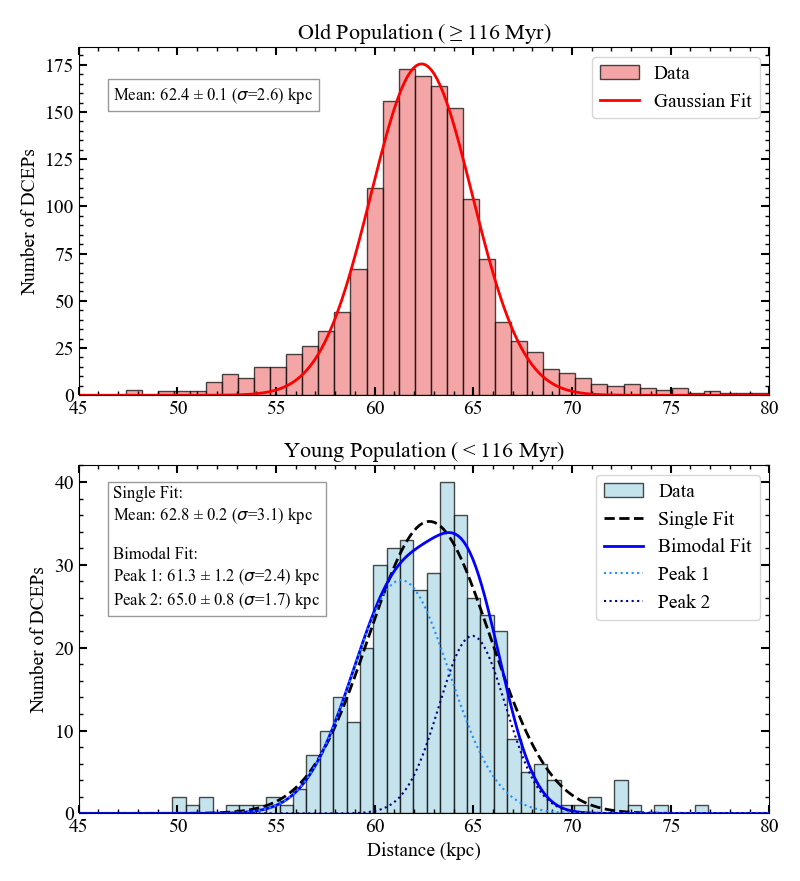} 
      \caption{Top panel: distance distributions for the "old" (older than 116 Myr) DCEPs in the SMC. The solid red line show the Gaussian fit to the histogram. The mean, its error and the dispersions are labelled. Bottom panel: as for the top panel but for the "young" population (younger than 116 Myr). The data have been fitted with single (black dashed line) and bimodal Gaussian functions (dotted lines, blue solid line). 
      The mean, its error and the dispersions for the different cases are labelled. 
      }
         \label{fig:relativeDistancesHistogram}
   \end{figure}

\section{Light curve analysis}

To put in phase the $J$ and $K_\mathrm{s}$ time series, we adopted the periods and epoch of maximum light calculated by the OGLE survey as mentioned in the previous section. 

A selection of the resulting phased light curves in $J$ and $K_\mathrm{s}$ at varying periods is shown in Figs.~\ref{fig:lc_LMC_J} to 
\ref{fig:lc_SMC_Ks} for the LMC and SMC, respectively. This atlas of light curves shows the change in shape of the light curves for varying modes, periods and (mean) metallicity, the latter being [Fe/H]$\sim-0.75$ dex and  $\sim-0.4$ dex, for the SMC and the LMC DCEPs, respectively \citep[][]{Romaniello2008,Romaniello2022}. 

To derive accurate PL and PW relations, we calculated average magnitudes directly from our light curves. Different from our previous papers \citep[e.g.][]{Ripepi2016,Ripepi2017,Ripepi2022b}, template-fitting is not required because our data is well-sampled with roughly 40 epochs per filter. We therefore modelled each star in both filters using a truncated Fourier series of the form $m(\phi)= A_0 + \sum^N_1 A_i\, {\rm cos}\,(2\pi i \phi + \Phi_i)$, 
where $N$ is the number of harmonics, $A_0$ is the average magnitude, $\phi$ is the phase (intended as the non-dimensional substitute of time; it is therefore an independent variable), while $A_i$ and $\Phi_i$ are the unknown coefficients of the series, i.e. the amplitude and the phase of each term.  
The fitting procedure started with a single harmonic component, and additional harmonics were iteratively included up to a maximum order determined through an F-test. The maximum number of harmonics was set to nine. However, for light curves characterised by a Uniformity Index ($UI$, see below) lower than 0.97, the maximum order was limited to six.\footnote{The adopted threshold values for both the maximum harmonic order and the Uniformity Index were determined empirically through dedicated tests performed on a representative subset of problematic light curves. The final values were selected based on a visual inspection of the resulting fits, aiming to minimise non-physical oscillations while preserving the intrinsic morphology of the light curves.} This choice was motivated by the need to prevent over-fitting effects, such as spurious oscillations, which become more prominent in sparsely and non-uniformly sampled phase distributions.

To further reduce these effects, phase gaps wider than 0.1 were filled by introducing one or two artificial data points obtained through linear interpolation between the observed measurements immediately preceding and following the gap. These interpolated points were used exclusively during the fitting procedure and were not considered in the subsequent analysis.

After obtaining an initial Fourier model, an outlier rejection procedure was applied. Residuals with respect to the fitted model were computed, and measurements with absolute residuals larger than three times the double Median Absolute Dispersion (MAD)\footnote{The double MAD is a robust estimator of dispersion on the left and right sides of the distribution median. This approach is particularly suitable for asymmetric residual distributions and is less sensitive to outliers than the standard deviation.} evaluated around the median were discarded. The Fourier fitting procedure was then repeated and applied to the cleaned light curve.

During this second fitting stage, bootstrap simulations were performed at each harmonic-order selection step to estimate the uncertainties associated with the fitted parameters. In particular, uncertainties were derived for the Fourier amplitudes and phases, the intensity-averaged mean magnitude, and the peak-to-peak amplitude. Bootstrap realisations were generated by resampling the light curve with replacement several times, equal to ten times the number of measurements in the original time series.\footnote{The adopted number of bootstrap realisations was selected as a compromise between computational efficiency and the statistical robustness of the resulting parameter distributions.} For each bootstrap realisation, the complete Fourier analysis was repeated, and the corresponding light curve parameters were measured. The uncertainty associated with each parameter was finally estimated from the resulting bootstrap distribution using the MAD.

As a result of this fitting technique, we obtained not only the average magnitude in both filters, but also the peak-to-peak amplitude of the pulsators (maximum$-$minimum magnitude of the fitting model), and the Fourier parameters $R_{21}$, $R_{31}$ and $\phi_{21}$,  $\phi_{31}$. These last quantities are fundamental tracers of light curve shape and will be discussed in detail below. 

In addition to the above-mentioned parameters, we have devised two indicators of the goodness of the light curve, and, consequently, of the quantities derived from them. 
The first is the above-mentioned $UI$, a parameter that evaluates the light curve's sampling. 
It was calculated according to the definition of \citet{madore2005nonuniform}:
 
\begin{equation}
    UI = \frac{N}{(N-1)} \Bigl[1 - \sum^N_{l=1} (\phi_{l+1}-\phi_l)^2 \Bigr],
\end{equation}

\noindent
where $N$ is the total number of epochs and $\phi$ is the phase. When interpreted as a normalised variance, the $UI$ statistic equals unity if the data points are uniformly and non-redundantly distributed across the light curve. Values of $UI>$0.95 indicate generally well-sampled light curves. In our sample, considering the LMC and SMC pulsators together, this happens in 96.5\% and 95\% of the cases in $J$ and $K_\mathrm{s}$, respectively. The average value is 0.975 in both filters. 

The second parameter is the light curve quality flag (lc\_qf hereafter), defined as the ratio between the standard deviation of the original time series and the rms of the residuals around the fitting line. This can be interpreted as a sort of signal-to-noise ratio. Values of lc\_qf$>$3 generally correspond to good quality light curves with low scatter. This is true in 85\% and 78.3\% of the cases, $J$ and $K_\mathrm{s}$, respectively. The average value of lc\_qf for the entire sample of DCEPs in the two bands is about 9 and 7.4.  

Therefore, our DCEP sample is overall composed of well-sampled light curves showing reduced scatter. 

The results of light curve fitting, including average magnitudes, amplitudes, Fourier parameters and uniformity/quality flags, are listed in Table~\ref{table:main_results}. All the following analyses are based on this table.

\section{Derivation of the PL and PW Relations}
\label{sect:pl}
In this section, we derive new PL and PW relations for the central regions of the LMC and SMC, based on the VMCDeep data described and analysed in the previous sections. These relations take the following mathematical form: 
\begin{equation}
    M_{\lambda}=a+b\times (\log P-\log P_0)  
\end{equation}
where $P_0$ is a pivoting period adopted to mitigate the correlation between $a$ and $b$. $M_{\lambda}$ can represent the magnitudes $J$, $K_\mathrm{s}$ or the Wesenheit magnitudes: $WJK_\mathrm{s}=K_\mathrm{s}-0.69\times(J-K_\mathrm{s})$ and  $WVK_\mathrm{s}=K_\mathrm{s}-0.13\times(V-K_\mathrm{s})$. Here, the coefficients have been chosen consistently with our previous papers and are therefore based on the \citet{Cardelli1989} extinction law (see Sect.~\ref{sect:reddening} for more details).

Before proceeding, we note that according to our previous works based on the VMC survey, some non-linearities are present in both the PL and PW relations in the LMC and SMC. Specifically, in \citet{Ripepi2017} we found that the PL and PW relations for F-mode pulsators in the SMC show a break at $P\sim2.95$ days. Similarly, in the LMC, a break was found by \citet{Ripepi2022b} for the 1O pulsators at $P\sim0.58$ days. Therefore, in the following, we will indicate with F\_a and F\_b
the solutions for fundamental pulsators in the SMC with periods shorter and longer than 2.95 days. Similarly, in the LMC, we indicated with 1O\_a and 1O\_b the first overtones with periods lower or larger than P=0.58 days. 

To derive the PL and PW relations, we employed a multi-tiered robust regression framework designed to mitigate the influence of photometric outliers and small-sample biases. 
The primary fitting technique relies on a Robust Linear Model (RLM) utilising M-estimators\footnote{M-estimators (Maximum likelihood-type estimators) are a broad class of robust statistical methods used to estimate parameters by minimising (or maximising) a specific objective function. They are a generalization of Maximum Likelihood Estimation (MLE) designed to reduce the disproportionate influence of outliers.}. Specifically, we adopted Tukey's Biweight objective function \citep{Tukey1960, Huber1981}, which dynamically down-weighs data points based on their residual distance from the median trend. If a point is close to the line, it gets a weight near 1.0. As a point moves farther away, its weight smoothly decreases.
If a point is exceptionally far away (an outlier), the Tukey algorithm assigns it a weight of exactly 0. Unlike standard sigma-clipping, which assigns binary weights (in or out), Tukey's Biweight smoothly reduces the influence of long-tailed noise, thereby providing highly stable zero points and slopes. 


For specific subsamples suffering from sparsity or severe selection effects—most notably the first overtone variables at short periods in the LMC ($1\mathrm{O\_a}$)—the unconstrained M-estimators can sometimes converge on unphysical local minima. To counteract this, we deployed an iterative Ordinary Least Squares (OLS) algorithm combined with a $4\sigma$ clipping threshold. Furthermore, to prevent the slopes from deviating unphysically due to the paucity of data points, we implemented bounded priors using the Trust Region Reflective algorithm \citep{Branch1999} via \texttt{scipy.optimize.lsq\_linear}. In this configuration, the slope $b$ is left free to vary, but strictly constrained within a tight interval ($\pm 0.1$) around established values from the literature \citep{Ripepi2017, Ripepi2022b}.

Finally, to address non-linearities in the PL and PW planes, split samples (e.g., LMC 1O and SMC Fundamental) were modelled using a continuous broken-line regression. Continuity across the pivot period ($P_{\rm split}$) was mathematically enforced by introducing a hinge function of the form $(b_2 - b_1) \times (\log P - \log P_{\rm split}) \times \Theta(P - P_{\rm split})$, ensuring a seamless transition between the two period regimes.

The number of pulsators adopted in this work is significantly smaller than that used in our VMC-based previous works \citep[especially in the LMC, see][]{Ripepi2017,Ripepi2022b}. This occurrence could lead to bias in the derived PL and PW relations. This is particularly expected in the case of a narrower period span, where the statistical leverage to securely anchor the slope is diminished.
To mitigate these limitations and provide an absolute comparative baseline, we additionally performed regressions where the slopes were fixed to canonical values established from extensive, well-characterised data sets \citep[][for the SMC and LMC, respectively]{Ripepi2017, Ripepi2022b}. 

Table~\ref{tab:pl_pw_results} presents the regression parameters for the PL and PW relations, calculated with both approaches, i.e. with the slopes treated as free parameters (left) or slopes fixed to literature values (right).
The fitting relations in the case of free slopes are shown on top of the data in Fig.~\ref{fig:PL_LMC} and ~\ref{fig:PL_SMC} for the LMC and SMC, respectively. Similar plots for the fixed slopes are shown in Fig.~\ref{fig:PL_LMC_fixed} and \ref{fig:PL_SMC_fixed}.

Comparing the two approaches reveals that fixing the slopes introduces only marginal increases in the rms scatter, indicating that the canonical slopes are highly compatible with our data. More importantly, fixing the slopes breaks the degeneracy between the slope and the zero-point ($a$). The zero-points extracted from the fixed-slope regressions are consequently more robust against localised sample biases, making them more reliable for deriving accurate relative distance moduli.

\section{Results}

In the following, we exploit the data presented in the previous sections. We compare our observations with up-to-date theoretical pulsational predictions; we adopt the same models to estimate the age of the DCEPs; finally, we use our precise PL and PW relations to estimate the  
relative distances between the MCs and use these measurements to constrain the metallicity dependence of the quoted relationships.

\subsection{Comparison of Theoretical Models with Observations}

From the Fourier decomposition of the observed light curves in the $J$ and $K_{\rm s}$ bands, we derived the amplitude ratios $ R_{k1}=\frac{A_k}{A_1}$, 
and the phase differences $\phi_{k1}=\phi_k-k\phi_1$
where $A_k$ and $\phi_k$ are the amplitude and phase of the $k$-th harmonic component, respectively. In particular, we considered the Fourier parameters $R_{21}$, $R_{31}$, and $\phi_{21}$, together with the peak-to-peak amplitudes in both passbands. These quantities were compared with the predictions of pulsation models.

For this purpose, we adopted the grid of pulsation models presented by \citet{Desomma2022}. Specifically, we considered the models calculated with two Mass--Luminosity (ML) relations, namely the canonical one neglecting core-overshooting, mass-loss and rotation effects \citep[Set A of][]{Desomma2022} and the mildly non-canonical one, assuming a moderate efficiency of core overshooting and/or mass loss and/or rotation \citep[Set B of][]{Desomma2022}. Moreover, two values of the mixing length parameter ($\alpha = 1.5$ and $\alpha = 1.7$) were adopted.

The theoretical bolometric light curves were transformed into the observational $J$ and $K_{\rm s}$ passbands using the bolometric correction tables provided by \citet{ChenYang2019}, which are based on PHOENIX atmosphere models. The resulting theoretical light curves were analysed using the same Fourier fitting procedure adopted for the observational data. From the corresponding Fourier coefficients, we derived the theoretical values of $R_{21}$, $R_{31}$, $\phi_{21}$, and the peak-to-peak amplitudes.

Figures~\ref{fig:fou_lmc} and ~\ref{fig:fou_lmc_a} show the comparison between observations for LMC DCEPs (Fourier parameters and peak-to-peak amplitude derived from $K_\mathrm{s}$-band light curves) and pulsation theory predictions based on models with $Z$=0.008 ML\,A and ML\,B, respectively. Left and right panels of each plot display model results for $\alpha$=1.5 and 1.7, respectively. The plots for the $J$ band show very similar results and are omitted for brevity. Figures~\ref{fig:fou_smc} and ~\ref{fig:fou_smc_a} show the same comparison, but for the SMC and correspondingly adopting models for $Z$=0.004.

These figures allow us a systematic qualitative comparison between the theoretical models and the observational data.
The comparison between model predictions and observations is mainly affected by the choice of the ML relation. Across both the LMC and SMC datasets, the ML\,B models provide a significantly better representation of the observational data than ML\,A. This is most clearly evidenced by the Fourier phase parameter, $\Phi_{21}$, at longer periods ($P > 10$ days). In the ML\,A models, the theoretical points diverge severely, dropping toward phase values between 2 and 4. In contrast, the observational data remain densely clustered at higher values, typically between 5 and 6. The ML\,B models accurately capture this morphological feature, gently curving to track the bulk of the observational points in this period regime.

Within the framework of the ML\,B relation, variations in the mixing length parameter yield more subtle, yet distinguishable, differences. A comparison of the amplitude parameters ($A_J$ and $A_{K_\mathrm{s}}$) reveals that the $\alpha = 1.5$ models systematically overestimate the maximum pulsation amplitudes at longer periods, reaching values upwards of 0.6. The $\alpha = 1.7$ models introduce a lower amplitude cap, peaking closer to roughly 0.4 to 0.5. This damping provides a much tighter overall fit to the observational constraints, restricting the extreme amplitude values predicted by the $\alpha = 1.5$ models and better aligning with the dense cores of the empirical distributions.

Based on the direct comparison with the empirical data, the theoretical model utilising the ML\,B relation coupled with a mixing length of $\alpha = 1.7$ emerges as the most accurate representation. This conclusion holds robustly across both galaxies (LMC and SMC) and both NIR photometric bands ($J$ and $K_\mathrm{s}$). From the theoretical point of view, the slightly non-canonical ML\,B can have different physical origins: overshooting in the main-sequence phase, mass--loss or rotation. It is very difficult to disentangle the contribution of each phenomenon. A detailed discussion of this point is beyond the scope of the present paper. The interested reader can find in-depth analysis in previous investigations which found very similar results to those presented above \citep[see e.g.][]{Marconi2017,Marconi2025,Molinaro2025}.

\subsection{Age of DCEPs in the MCs centres}
\label{sect:age}
The existence of Period-Age (PA) and Period-Age-Colour (PAC) relations for DCEPs 
makes these pulsators important tracers of the young stellar populations in the host galaxies. In our previous works, we derived the ages of almost all DCEP samples in both MCs. In the LMC, \citet{Ripepi2022b} found a main burst of DCEPs formation at about 93 Myr ($\sigma\sim21$ Myr), and a much smaller one, at 159 Myr ($\sigma\sim17$ Myr). Instead, in the SMC \citet{Ripepi2017} found that the age distribution was bimodal, with two peaks at 120 and 220 Myr ($\sigma\sim10$ Myr for both peaks) and the old DCEPs more spatially elongated with respect to the young ones, which appeared more centrally concentrated. 
It is interesting to investigate the ages of the DCEPs in the central regions of LMC and SMC based on the new data presented in this paper. To this aim, we adopted the PAC relations calculated by \citet{Desomma2021} for $Z$=0.008 and $Z$=0.004 for the LMC and SMC, respectively\footnote{We note that in this case we adopted the "Set A" models, because the ages for 1O pulsators were not available for "Set B" models in \citet{Desomma2021}. This choice allows us to calculate consistent ages between different pulsation modes.}. As for the colour, we adopted the (dereddened) $(J-K_\mathrm{s})$, so that we use only the highly precise data from VMCDeep. The ages of the targets and their uncertainties, calculated propagating the errors on the coefficients of the PAC relation and on the observed $(J-K_\mathrm{s})$, are listed in Table~\ref{table:main_results}. Figure~\ref{fig:ageHistograms} displays the age distributions for both the LMC and SMC. The result for the LMC is almost identical to that obtained in our previous work by using the whole DCEPs sample. Indeed, we observe a peak at 92.0$\pm$1.3 Myrs (dispersion 22.8 Myr, see appendix~\ref{app:error} for details on the calculation of the errors on the mean) and a hint of some enhanced formation at ages later than 150 Myrs, not precisely measurable in the present data due to the poor statistics. 
This result is not surprising as we used the same models as in \citet{Ripepi2022b}, albeit with a different sample and different colours. 
More interesting is the case of the SMC, where the bi-modality of the DCEPs populations is confirmed, with a smaller peak at 91.5$\pm$4.0 Myr ($\sigma\sim27.9$ Myr), and a larger one at 187.6$\pm$3.8 Myr ($\sigma\sim48$ Myr). The marginally significant difference with \citet{Ripepi2017} may originate from the use of a different sample of DCEPs considered here, and/or by the adoption of older pulsational models \citep[e.g.]{Bono2005} in our previous work.

It is crucial to emphasise that this new analysis, which targets the galactically similar central regions of both Clouds, reveals that the young DCEP populations in the SMC and LMC are not merely similar in age, but exactly co-eval within the uncertainties. While the intrinsic age dispersion of these star formation episodes is significant ($22.8$~Myr for the LMC and $27.9$~Myr for the SMC, our rigorous statistical analysis demonstrates that the peaks of the distributions are determined to a much higher precision: $1.3$~Myr for the LMC and $4.0$~Myr for the SMC's younger peak. This strict synchronicity ($92.0 \pm 1.3$~Myr vs $91.5 \pm 4.0$~Myr) suggests that a major, simultaneous event triggered the star formation in both systems---most likely a recent direct collision between the LMC and SMC. Furthermore, our findings highlight an asymmetry of the older DCEP population. 
The SMC exhibits a prominent secondary star formation peak at $187.6 \pm 3.8$~Myr ($\sigma=48.0$~Myr), which is virtually absent in the LMC. This may imply that an earlier interaction must have profoundly affected the SMC while leaving the LMC largely undisturbed, pointing towards tidal effects stemming from a close, but not too close, earlier encounter. These observationally derived, dynamically driven star formation epochs align well with recent kinematic evidence and $N$-body hydrodynamical simulations of the Magellanic system, which predict a series of close encounters concluding with a recent direct collision between the Clouds within the last $\sim 100-200$~Myr \citep[see e.g.,][and references therein]{Besla2016, Zivick2018, Choi2022, Ficara2026}.

\subsubsection{Spatial distribution of the young and old DCEP populations in the SMC}

It is interesting to analyse the spatial distribution of the two age--separated populations in the SMC to investigate whether the two recent episodes of star formation are spatially coincident. Spatial differences could be linked to strong tidal interactions between the SMC and LMC, which are likely responsible for the star formation bursts that led to the two peaks in the DCEP age distribution.

For this purpose, we divided the DCEP sample into ``young'' and ``old'' populations based on the bimodal Gaussian profile fit to their age distribution. The intersection of the two Gaussian components at $116$~Myr provides a natural boundary, yielding 410 young DCEPs ($t < 116$~Myr) and 1519 old DCEPs ($t \ge 116$~Myr). 
Next, we computed the relative distance moduli ($\Delta \mu$) for all variables in both age groups by comparing them against the empirical PWJKs\footnote{Period--Wesenheit relation in the $J,K_\mathrm{s}$} ridge lines for the F\_a, F\_b, and 1O modes (detailed in Table~\ref{tab:pl_pw_results}). Finally, individual physical distances were estimated via the relation $D (\mathrm{kpc}) = 62.4 \times 10^{\Delta \mu / 5}$. Here, 62.4 kpc represents the adopted mean distance to the SMC \citep{Graczyk2020}; however, we note that the exact zero-point of this distance is inconsequential for examining the relative spatial differences between the two populations.

The distance distribution of the two DCEP populations reveals a striking structural complexity. The older DCEP population ($\ge 116$~Myr) exhibits a relatively smooth, single-peaked spatial distribution. It is well-described by a single Gaussian peaking at $62.4 \pm 0.1$~kpc with a line-of-sight dispersion of $\sigma = 2.6$~kpc. This value perfectly traces the established geometric centre and the considerable structural depth of the SMC.

In contrast, the younger population ($< 116$~Myr) shows clear evidence of a bimodal distribution along the line of sight. We fitted this younger population with a bimodal Gaussian, using robust bootstrap resampling to estimate the errors on the mean of the two peaks. This procedure confirms that the two peaks are highly significant and physically distinct, with means securely separated at $61.3 \pm 1.1$~kpc ($\sigma=2.4$~kpc) and $65.0 \pm 0.7$~kpc ($\sigma=1.7$~kpc). The spatial separation of almost $4$~kpc is far larger than the combined statistical uncertainties on the respective means. 



A comparable line-of-sight dichotomy among young, massive stars was identified by \citet{Murray2024}. Using a sample of young (10–100 Myr) Red Supergiant (RSG) stars and neutral hydrogen (HI) kinematics, they modelled the SMC as two distinct star-forming systems separated by approximately 5 kpc (at distances of $\sim$61 and $\sim$66 kpc). 
This scenario aligns with a long-standing history of observational evidence for line-of-sight substructures and distance bimodality in the SMC, spanning early stellar and gaseous kinematics \citep[e.g.][]{Mathewson1984, Hatz1989} to modern distance determinations using intermediate-age population tracers such as RC stars \citep[][]{Nidever2013, Subramanian2013}.
Our Cepheid distance measurements provide a highly precise, independent standard-candle confirmation of this ``two-sheet'' structure. To rule out the possibility that our observed separation is a photometric artefact caused by differential extinction---where a heavily obscured population appears artificially fainter and therefore more distant---we divided the young sample into a near ($D < 62$ kpc) and far ($D > 63$ kpc) component. We found almost identical values for the mean $E(V - I)$ reddening between the two groups ($\sim$0.065 mag), confirming that the spatial separation is a genuine geometric feature.

This bolsters the scenario that a recent, strong tidal interaction---likely a direct collision with the LMC---has not only triggered a synchronised burst of star formation, but has also distinctly disrupted the SMC's gaseous and young stellar components along the radial direction.

Interestingly, a distance dichotomy by 2-2.5 kpc in the central region has also been found in the distribution of RC stars \citep[][]{Omkumar2026}, based on the same VMCDeep data. However, it is not yet clear if the two structures are related. If they refer to the same structures along the line of sight, it could be that the DCEPs formed later and are located at a greater distance from each other, given that the gas is already expanding and moving at a certain velocity (as HI studies indicate). This means either that the moment of the collision also influenced the pre-existing stars—which, in any case, move more slowly than the gas—or that the RC stars and DCEPs trace interactions that occurred at different times. 
The future availability of accurate proper motions from the $Gaia$ mission Data Release 4 and, mainly, radial velocity for hundreds of thousands of stars in the MCs thanks to the soon-to-start "The One Thousand and One Magellanic Fields (1001MC) Survey" \citep[][]{Cioni2019}, will provide the key to fully explain the above-discussed findings. 

\subsection{Relative distance between the LMC and SMC: a test for the metallicity dependence of PL and PW relations}
\label{sect:relativeDistance}

The availability of precise PL and PW relations for the central fields of the LMC and SMC allows us to carry out an estimate of the relative distance between the centres of the two galaxies. This has already been done in the past without reaching consensus on a unique set of values, as shown in Table~\ref{tab:lmc_smc_distance}. However, to our knowledge, no previous work adopted a correction for the different average metallicity between the two Clouds, which is about $\Delta$[Fe/H]$\sim$0.34 dex \citep{Romaniello2008,Romaniello2022}.

\begin{table}
\centering
\caption{Literature relative distance estimates between the LMC and SMC using DCEPs}
\label{tab:lmc_smc_distance}
\footnotesize\setlength{\tabcolsep}{3pt}
\begin{tabular}{l c l }
\toprule
Study & $\Delta\mu$\,(SMC$-$LMC) & Method/Bands \\
      & (mag)                  &                 \\
\midrule
\citet{degrijs2014}\\
\citet{degrijs2015}& $0.50 \pm 0.15$ & Mix \\
\citet{Ngeow2015}& $0.483 \pm 0.015$ & PW (V,I,J,H,K \& IRAC) \\
\citet{Ripepi2016}   & $0.55 \pm 0.04$   & PW NIR (VMC survey)\\
\citet{Scowcroft2016} & $0.48 \pm 0.01$   & PL Mid-IR (IRAC)  \\
\bottomrule
\end{tabular}
\tablefoot{VMC is the VISTA Magellanic Cloud survey defined in the introduction; IRAC is the Infrared Array Camera (IRAC) on the Spitzer Space Telescope, which operated in the mid-infrared spectrum.}
\end{table}

The precision of the present data stimulated us to carry out such metallicity-dependent relative distance calculation, which, at the same time, can be used to test the metallicity dependence of PL and PW relations, a still highly uncertain and debated topic in the recent literature. Indeed, while the SH0ES \citep[Supernovae, H0, for the Equation of State of Dark energy; ][]{Riess2016} 
team finds metallicity terms ($\gamma$) of the order of $-0.2$ mag dex$^{-1}$ \citep[see][and references therein]{Riess2022a,breuval2024}, the C-MetaLL survey \citep[Cepheid—Metallicity in the Leavitt Law;][]{Ripepi2021a}, in a series of papers, consistently derived values of $\gamma\leq-0.4$ mag dex$^{-1}$ \citep[][]{Trentin2024_IV,Ripepi2026}\footnote{The C-MetaLL results have been replicated by independent researchers, e.g. \citet{WangChen2025}.}. Although this discrepancy has only a minimal impact on the $H_0$ tension \citep[][]{Riess2022a}, nevertheless, knowing with precision the value of $\gamma$ is important in many aspects \citep[see discussion in e.g.][]{Ripepi2026}.

The metallicity dependence of the PL and PW relations presented here was tested as follows. First, we calculated the relative distance modulus between the SMC and LMC ($\Delta\mu = \mu_{\text{SMC}} - \mu_{\text{LMC}}$) across multiple bands and fitting models. Then, we compared these empirical photometric distances against the highly precise, geometric distance difference established by late-type detached eclipsing binaries (DEBs). Adopting $\mu_{\text{LMC}} = 18.477 \pm 0.026$ mag \citep{Pietrzynski2019} and $\mu_{\text{SMC}} = 18.977 \pm 0.030$ mag \citep[]{Graczyk2020}, the geometric anchor is $\Delta\mu_{\text{bin}} = 0.500 \pm 0.040$ mag.
This apparently simple procedure is described in detail in the next sections.

\begin{figure}[h!]
    \centering
    \includegraphics[width=8.5cm]{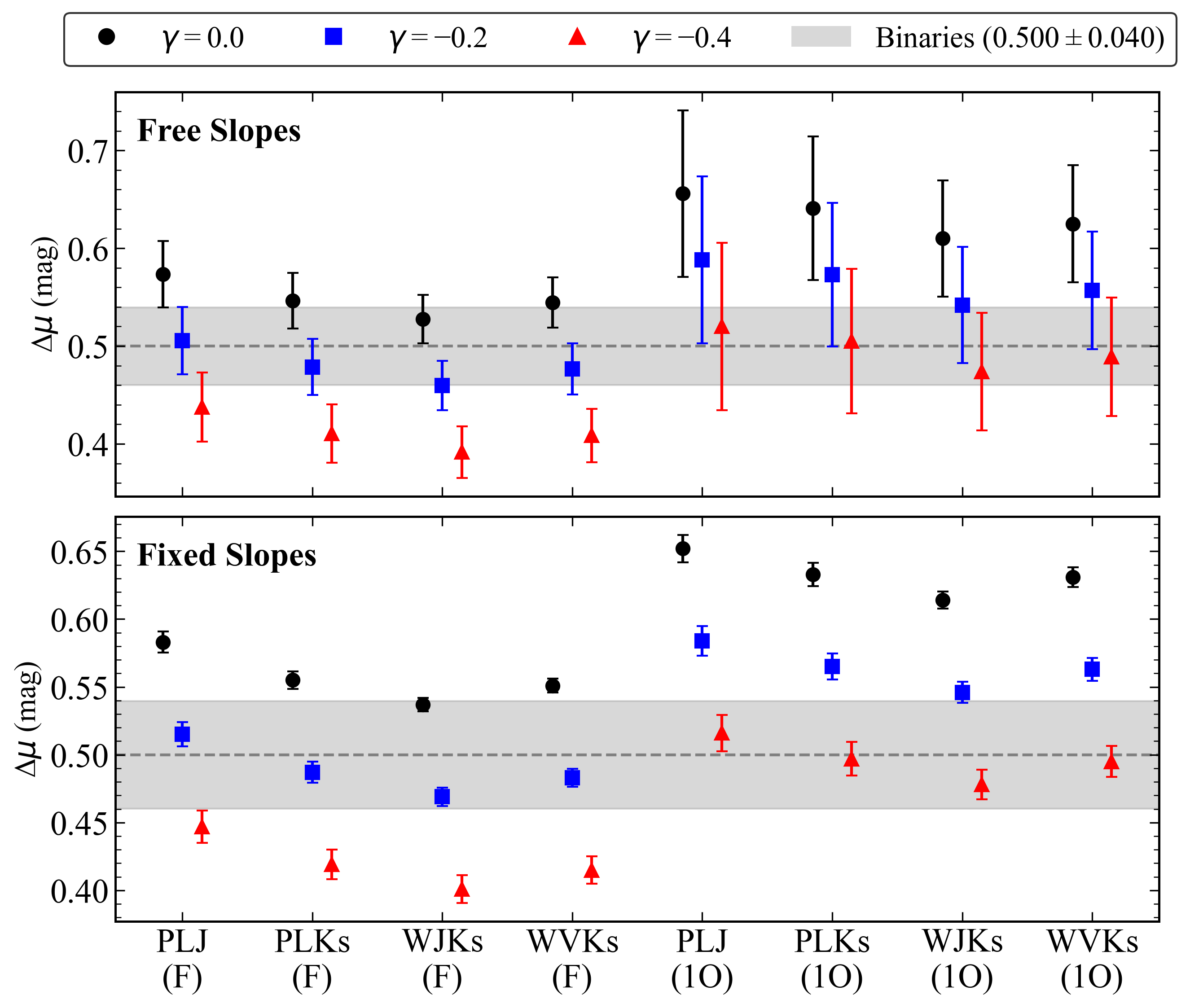}
    \caption{Relative distance moduli between the SMC and LMC for the separated F and 1O samples, evaluated across different photometric bands and metallicity assumptions ($\gamma = 0.0, -0.2, -0.4$ mag dex$^{-1}$). The upper panel displays the free-slope paradigm, and the lower panel displays the fixed-slope paradigm. The horizontal shaded band represents the geometric evaluation $\Delta\mu = 0.500 \pm 0.040$ mag established by means of the DEBs.}
    \label{fig:separated_modes}
\end{figure}

\begin{figure}[h!]
    \centering
    \includegraphics[width=9cm]{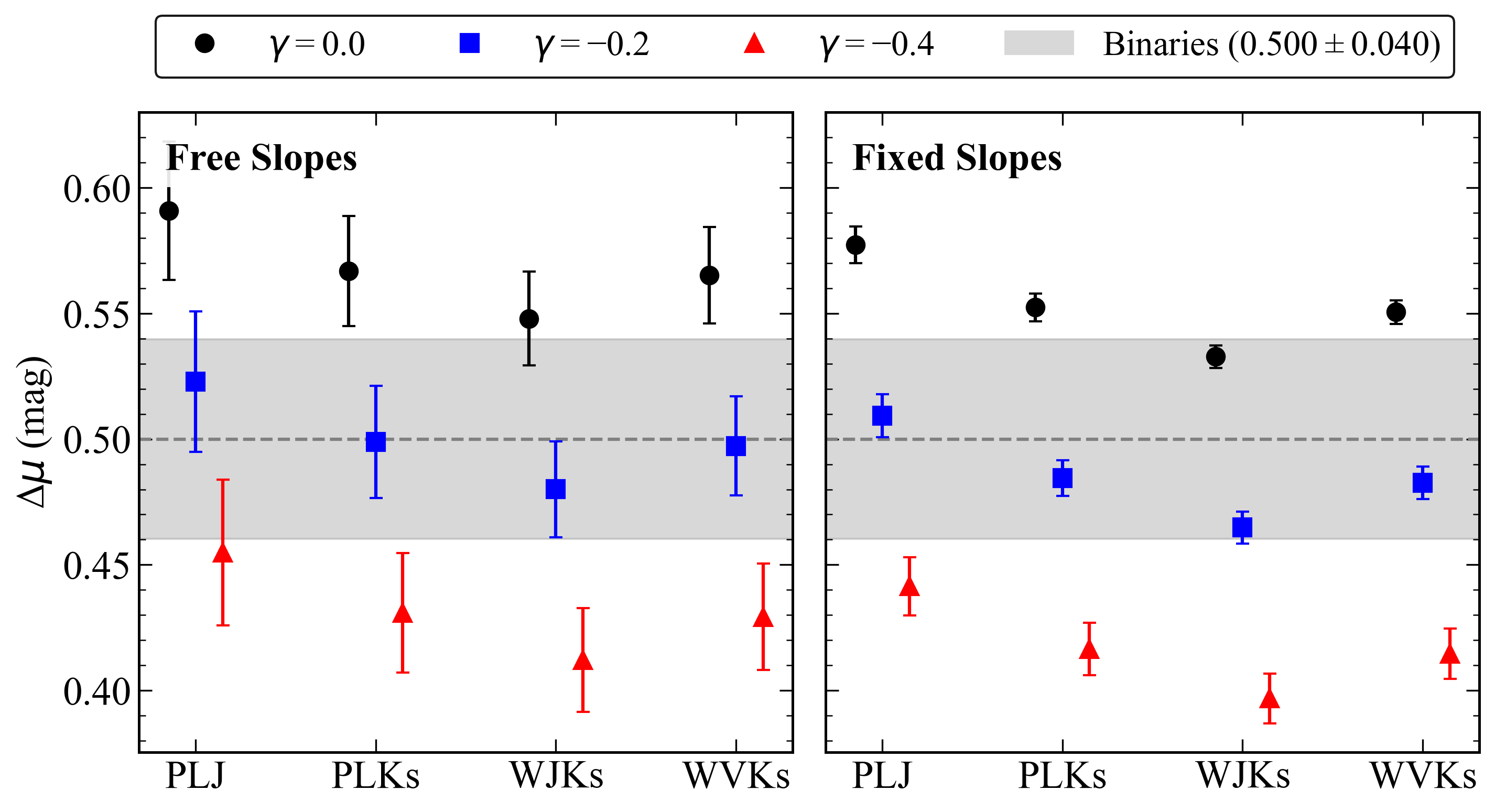}
    \caption{Relative distance moduli between the SMC and LMC derived from the combined, fundamentalised sample. The left panel shows the free-slope fits, and the right panel shows the fixed-slope fits. Reddening-free Wesenheit relations ($W_{J,K_\mathrm{s}}$ and $W_{V,K_\mathrm{s}}$) with a moderate metallicity correction ($\gamma = -0.2$ mag dex$^{-1}$) agree best with the geometric binary distance (horizontal shaded band).}
    \label{fig:fundamentalized_modes}
\end{figure}

\subsubsection{Calculation of Relative Distances and Error Propagation}

Because the slopes of the PL and PW relations differ between the LMC and SMC (when left as free parameters), the calculated relative distance intrinsically depends on the period used to compare the two galaxies. Extrapolating the relations over long baselines can introduce severe, unphysical biases. To resolve this degeneracy, we evaluated the apparent magnitudes at a reference period ($\log P_{\text{ref}}$) located near the barycentre of the respective period distributions, where the statistical uncertainty of the regression is minimised. We adopted $\log P_{\text{ref}} = 0.5$ ($P \approx 3.16$ days) for F DCEPs and $\log P_{\text{ref}} = 0.0$ ($P = 1$ day) 1O pulsators.

The statistical error on the apparent magnitude at the reference period was propagated from the regression covariance matrix. Assuming minimal covariance between the slope ($b$) and the intercept evaluated at the pivot period ($a_{\text{pivot}}$), the magnitude error was:
\begin{equation}
    \sigma_{m}^2 = \sigma_{a_{\text{pivot}}}^2 + (\log P_{\text{ref}} - \log P_{\text{pivot}})^2 \sigma_b^2
\end{equation}
The raw relative distance modulus and its associated error were thus $\Delta\mu_{\text{raw}} = m_{\text{SMC}} - m_{\text{LMC}}$ and $\sigma_{\Delta\mu_{\text{raw}}} = \sqrt{\sigma_{m,\text{SMC}}^2 + \sigma_{m,\text{LMC}}^2}$.

To account for the metallicity differences between the two environments, we adopted mean spectroscopic abundances of $\text{[Fe/H]}_{\text{LMC}} = -0.41 \pm 0.01$ dex and $\text{[Fe/H]}_{\text{SMC}} = -0.75 \pm 0.02$ dex, yielding $\Delta\text{[Fe/H]} = -0.34 \pm 0.02$ dex. The metallicity-corrected relative distance modulus is defined as:
\begin{equation}
    \Delta\mu_{\text{corr}} = \Delta\mu_{\text{raw}} - \gamma \Delta\text{[Fe/H]}
\end{equation}
where $\gamma$ is the metallicity term in mag dex$^{-1}$. The final error on the corrected distance modulus accurately propagates the uncertainties in both the photometric fits and the spectroscopic metallicities:
\begin{equation}
    \sigma_{\Delta\mu_{\text{corr}}} = \sqrt{\sigma_{\Delta\mu_{\text{raw}}}^2 + \gamma^2 \left(\sigma_{\text{[Fe/H]},\text{SMC}}^2 + \sigma_{\text{[Fe/H]},\text{LMC}}^2\right)}
\end{equation}

\subsubsection{Fundamentalisation of first overtones}

Initial evaluations of $\Delta\mu_{\text{corr}}$ revealed a systematic discrepancy between the distances derived from F and 1O pulsators, with 1O DCEPs consistently yielding longer relative distances (by $\sim 0.08 - 0.10$ mag). This diversity arises from the intrinsic differences in their evolutionary crossing timescales, temperature boundaries within the instability strip, and mass distributions \citep[see e.g.,][]{Bono2000, Marconi2005}. Furthermore, 1O DCEPs had significantly narrower period distributions \citep{Sos2015}, making their unconstrained regression slopes highly susceptible to sample-specific geometric depth effects—particularly within the highly elongated structure of the SMC \citep{Jacy2016, Ripepi2017}—and statistical noise.

To maximise statistical leverage and eliminate the F/1O dichotomy, we applied a ``fundamentalisation'' procedure. The 1O variables were mapped onto the Fundamental PL and PW planes using the empirical period--ratio relations derived by \citet{Pilecki2024}. Specifically, the periods were shifted using their eq. 1b for the LMC:
\begin{equation}
    P_F / P_{\rm 1O} = 1.367 + 0.079 \log P_{\rm 1O}
\end{equation}
and eq. 1c for the SMC:
\begin{equation}
    P_F / P_{\rm 1O} = 1.356 + 0.068 \log P_{\rm 1O}
\end{equation}
Merging the fundamentalised 1O variables with the F variables yielded a single, highly constrained relative distance measurement for each photometric band. The results for the F, 1O, and fundamentalised (combined) samples are summarised in Table~\ref{tab:relative_distances}.

\subsubsection{Constraints on the Metallicity Term}

To visually assess the validity of the adopted metallicity terms, we plot the derived $\Delta\mu$ against the geometric DEB target band ($0.500 \pm 0.040$ mag). 

Figure \ref{fig:separated_modes} illustrates the results for the unmerged F and 1O samples. The significant vertical offset between the results with the two modes underlines the necessity of fundamentalising the 1O pulsators. The uncorrected ($\gamma = 0.0$ mag dex$^{-1}$) distances are systematically too large across almost all bands, demonstrating that a metallicity correction is required to reconcile the photometric and geometric distance scales.

Figure~\ref{fig:fundamentalized_modes} displays the highly constrained results after fundamentalising the 1O pulsators. By evaluating these combined samples, particularly using the PWJKs and PWVKs\footnote{Period--Wesenheit relation in the $V,K_\mathrm{s}$} relations, 
it becomes evident that a significant metallicity dependence is required. In the fixed-slope paradigm, which shows much smaller errors, a correction of $\gamma = -0.2$ mag dex$^{-1}$ shows the best agreement with the DEB measurements. In the debate about the actual size of the $\gamma$ value, this result seems to favour a mild metallicity dependence of the DCEPs PL and PW relation \citep[see e.g.][and references therein, for a discussion and a summary of the latest results]{Ripepi2026}.

\section{Conclusions}

In this paper, we have utilised deep NIR time series data from the VMCDeep survey to perform a detailed analysis of DCEP populations within the highly crowded and dusty central $1.77\text{ deg}^2$ regions of the LMC and SMC. By cross-matching our NIR catalogues with optical data from the OGLE survey, we secured a final sample of 573 DCEPs in the LMC and 2090 in the SMC. High-quality $J$ and $K_\mathrm{s}$ light curves, sampled with a nominal 40 epochs per filter, were directly modelled using a truncated Fourier series with robust outlier rejection and bootstrap error estimation.
Based on these data, the main results of this work are:

\begin{itemize}

\item Light curve atlas:
We produced an atlas of high-precision light curves at varying periods in $J$ and $K_\mathrm{s}$ bands for both the LMC and SMC. This atlas can serve as a reference for NIR light curves at the present-day typical metallicities of the two galaxies, namely [Fe/H]$\sim-0.4$ dex and $\sim-0.75$ dex for LMC and SMC, respectively.

\item PL and PW Relations: 
We derived the most precise and least dispersed near-infrared PL and PW relations ever published for the central regions of the Magellanic Clouds. By anchoring our slopes to well-characterised global samples, we overcame local period-span limitations, yielding zero-points of exceptional stability. These highly constrained relations provide a remarkably robust anchor for the cosmic distance ladder.     

\item Refined theoretical constraints:
We exploited the accurate Fourier parameters determined from light curve fitting by comparing our empirical findings with modern theoretical pulsation models. This comparison revealed that the ML relation "Set B" \citep[][]{Desomma2022} coupled with a mixing length parameter of $\alpha = 1.7$ yields the most accurate representation of the data. This configuration successfully tracks the morphology of the Fourier phase parameter $\Phi_{21}$ at long periods ($P > 10\text{ days}$) and correctly bounds the peak pulsation amplitudes.

 \item A synchronised, disruptive collision: Using Period-Age-Colour relations, we uncovered a strict synchronicity in the youngest star formation episodes of both Clouds, peaking simultaneously at $t \approx 92$~Myr. 
Crucially, our data provide highly precise, independent confirmation of a striking line-of-sight spatial dichotomy within this young SMC population. While the older SMC Cepheids ($\geq 116$ Myr) form a cohesive, single-peaked structure, the young population is cleanly split into two distinct spatial ``sheets'' separated by nearly 4 kpc. We verified through reddening analysis that this is a genuine geometric feature rather than a dust artefact. This perfectly corroborates the findings based on RSGs and HI gas \citep{Murray2024}, supporting the scenario of a disrupted, two-sheet SMC resulting from a recent direct collision with the LMC.
 

\item Mild PL and PW metallicity dependence: 
We used our PL and PW relations to calculate the relative distance between the MCs. We observed that standing alone, 1O DCEPs systematically yield longer relative distance moduli than F pulsators by $\sim 0.08$–$0.10\text{ mag}$. Applying an empirical "fundamentalisation" procedure effectively eliminates this dichotomy, unlocking a unified, highly constrained relative distance dataset.

Comparing our photometric relative distances against the precise geometric baseline provided by detached eclipsing binaries ($\Delta\mu_{\rm bin} = 0.500 \pm 0.040\text{ mag}$) confirms that a metallicity correction is mandatory, as uncorrected frameworks ($\gamma = 0.0\text{ mag dex$^{-1}$}$) systematically overestimate the distance scale. Adopting the fixed-slope PL and PW relations, a moderate metallicity term of $\gamma = -0.2\text{ mag dex$^{-1}$}$ combined with reddening-free Wesenheit relations ($W_{J,K_\mathrm{s}}$ and $W_{V,K_\mathrm{s}}$) offers the most robust and stable agreement with the geometric anchor.
\end{itemize}

\section{Data availability}
Table~\ref{tab:timeSeries}, including epoch PSF photometry for all the objects used in this work, is only available in electronic form at the CDS via anonymous ftp to cdsarc.u-strasbg.fr (130.79.128.5) or via http://cdsweb.u-strasbg.fr/cgi-bin/qcat?J/A+A/.

\begin{acknowledgements}
We thank our anonymous referee for providing helpful comments.
We acknowledge funding from: INAF GO-GTO grant 2023 “C-MetaLL - Cepheid metallicity in the Leavitt law” (P.I. V. Ripepi); 
PRIN MUR 2022 project (code 2022ARWP9C) 'Early Formation and Evolution of Bulge and Halo (EFEBHO),' PI: Marconi, M., funded by the European Union – Next Generation EU; Large Grant INAF 2023 MOVIE (P.I. M. Marconi).
This research has made use of the SIMBAD database operated at CDS, Strasbourg, France.
GDS acknowledges funding from Gaia DPAC through INAF and ASI (2025-10-HH.0; PI: M.G. Lattanzi) and INFN, Naples section, for support via QGSKY and Moonlight2 initiatives.
\end{acknowledgements}

\bibliographystyle{aa} 
\bibliography{mybib} 

\onecolumn
 \begin{appendix}

\section{Atlas of light curves}

    \begin{figure*}
    \centering
    \includegraphics[angle=0,width=18cm]{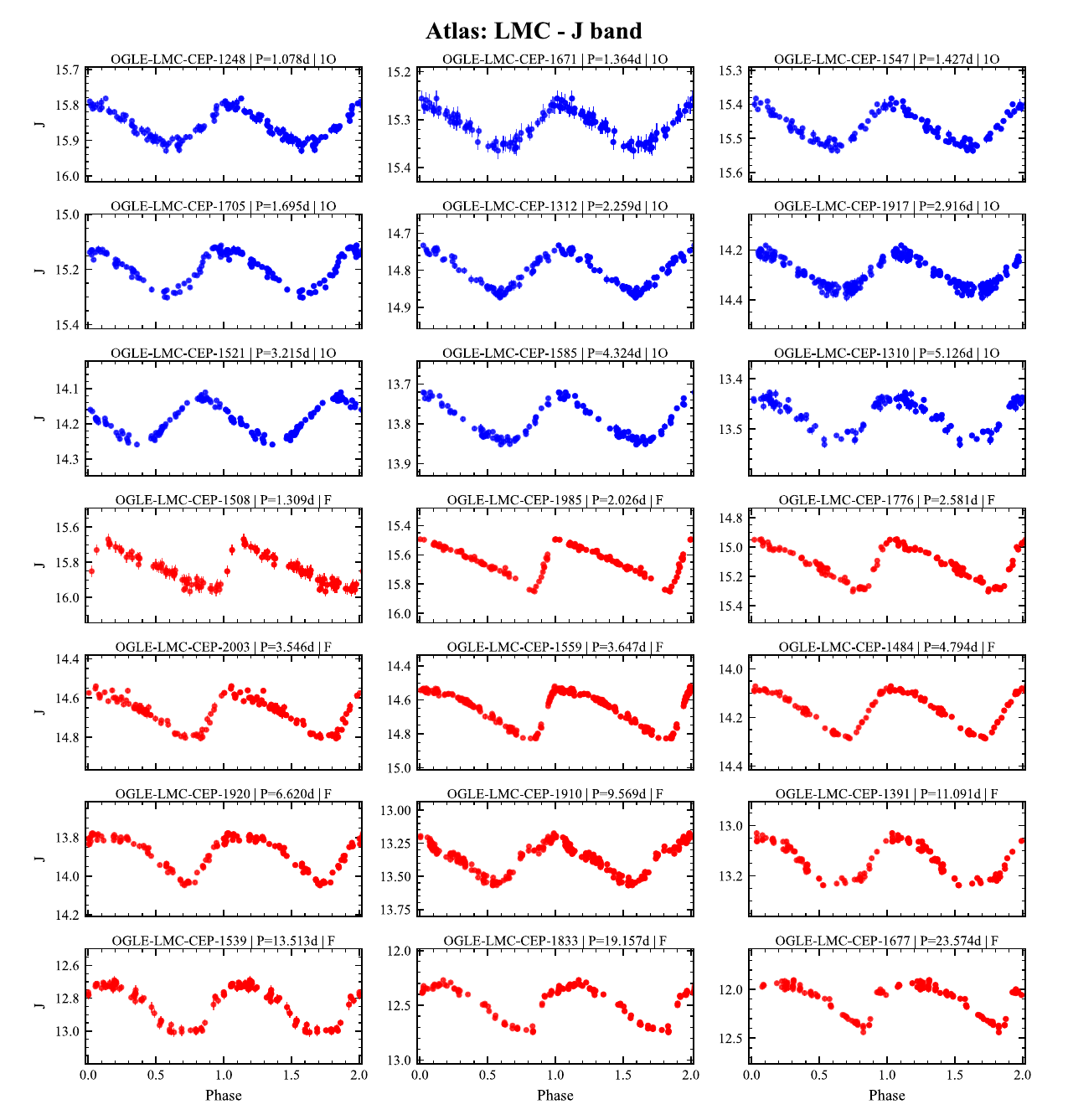}
       \caption{Atlas of light curves in the $J$ band at increasing periods for 1O and F pulsators (blue and red dots, respectively) in the LMC.}
          \label{fig:lc_LMC_J}
    \end{figure*}

    \begin{figure*}
    \centering
    \includegraphics[angle=0,width=18cm]{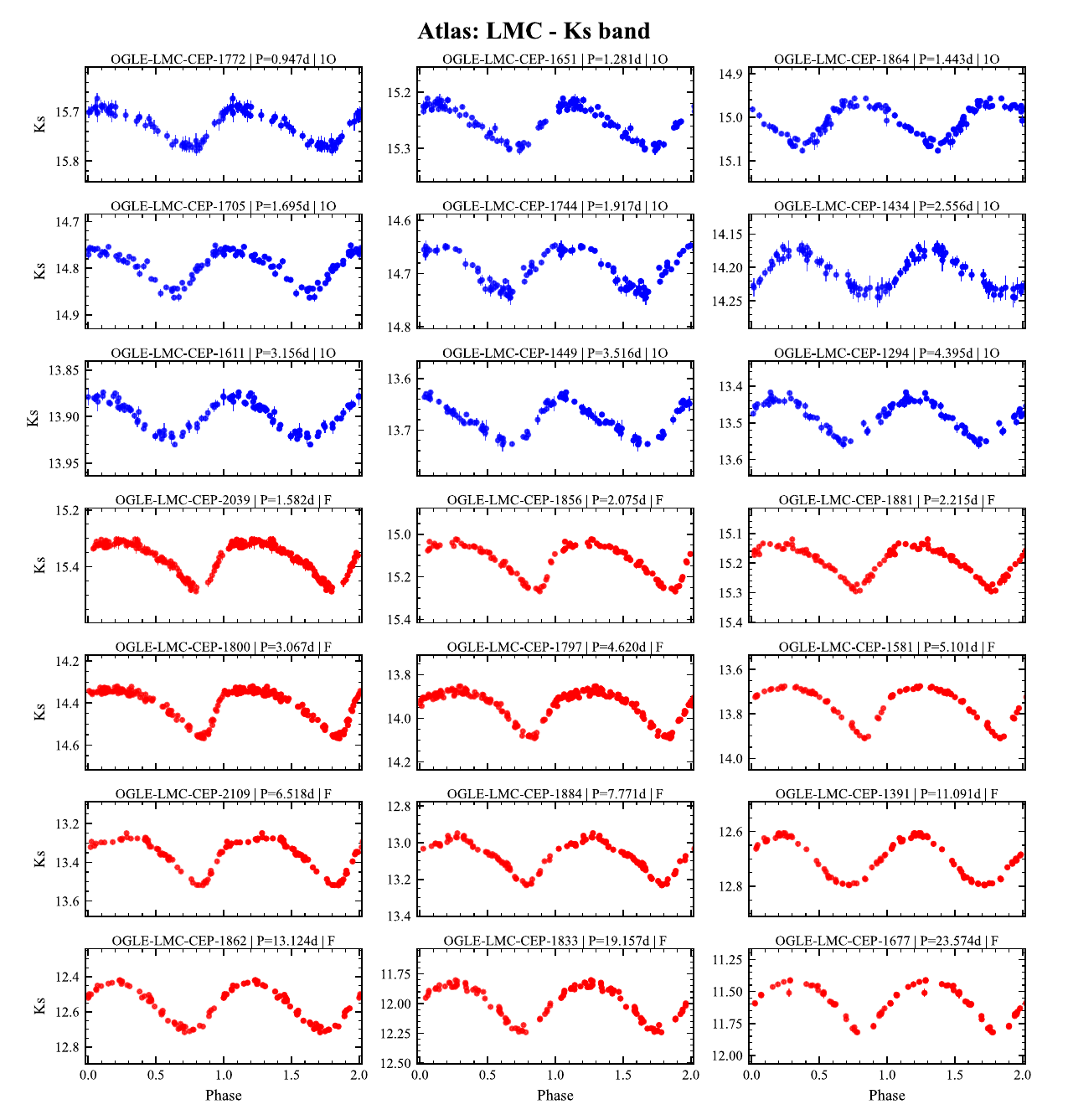}
       \caption{As in Fig.~\ref{fig:lc_LMC_J}, but for the $K_\mathrm{s}$ band.}
          \label{fig:lc_LMC_Ks}
    \end{figure*}

        \begin{figure*}
    \centering
    \includegraphics[angle=0,width=18cm]{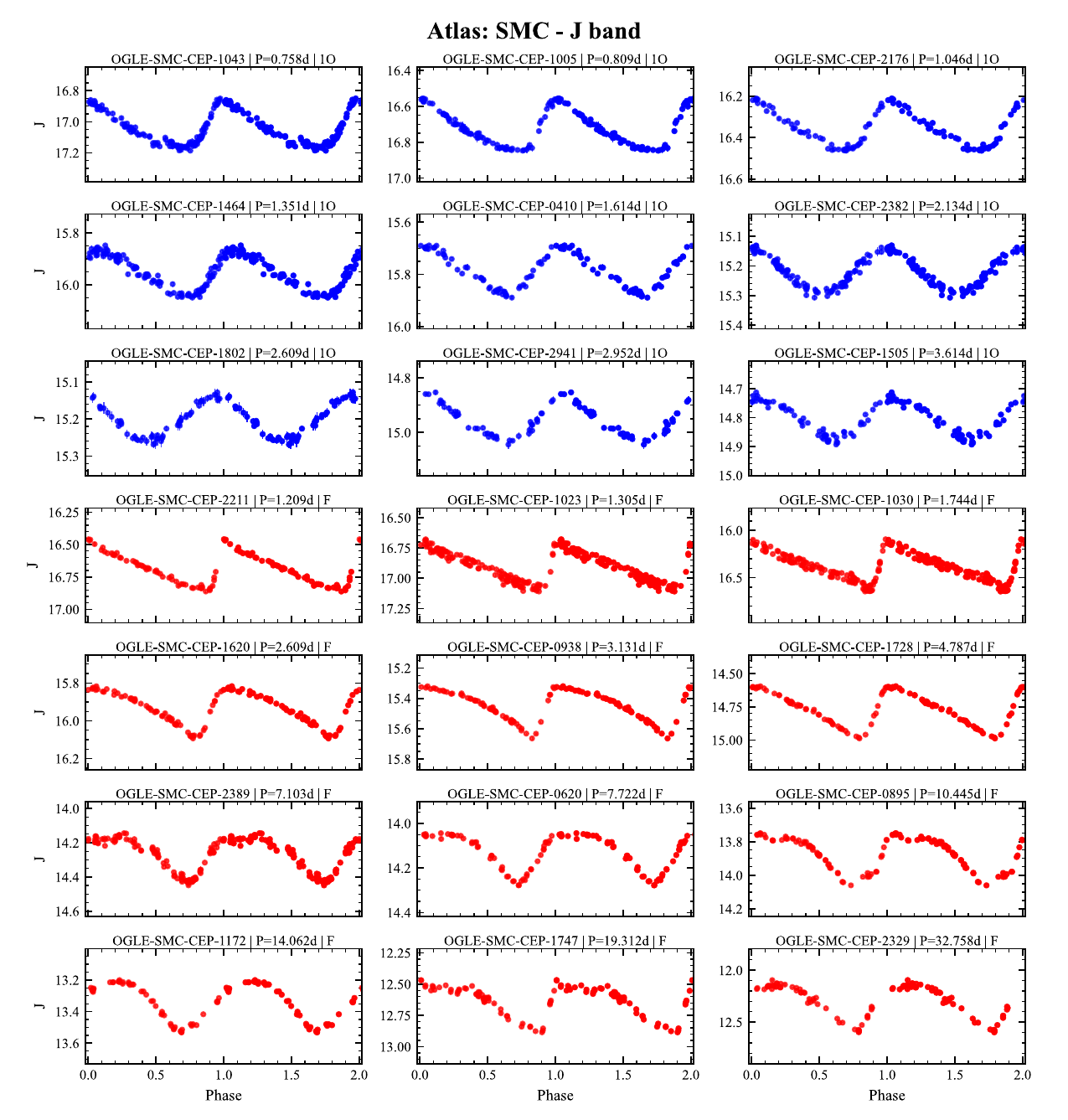}
       \caption{As in Fig.~\ref{fig:lc_LMC_J}, but for the SMC.}
          \label{fig:lc_SMC_J}
    \end{figure*}

       \begin{figure*}
    \centering
    \includegraphics[angle=0,width=18cm]{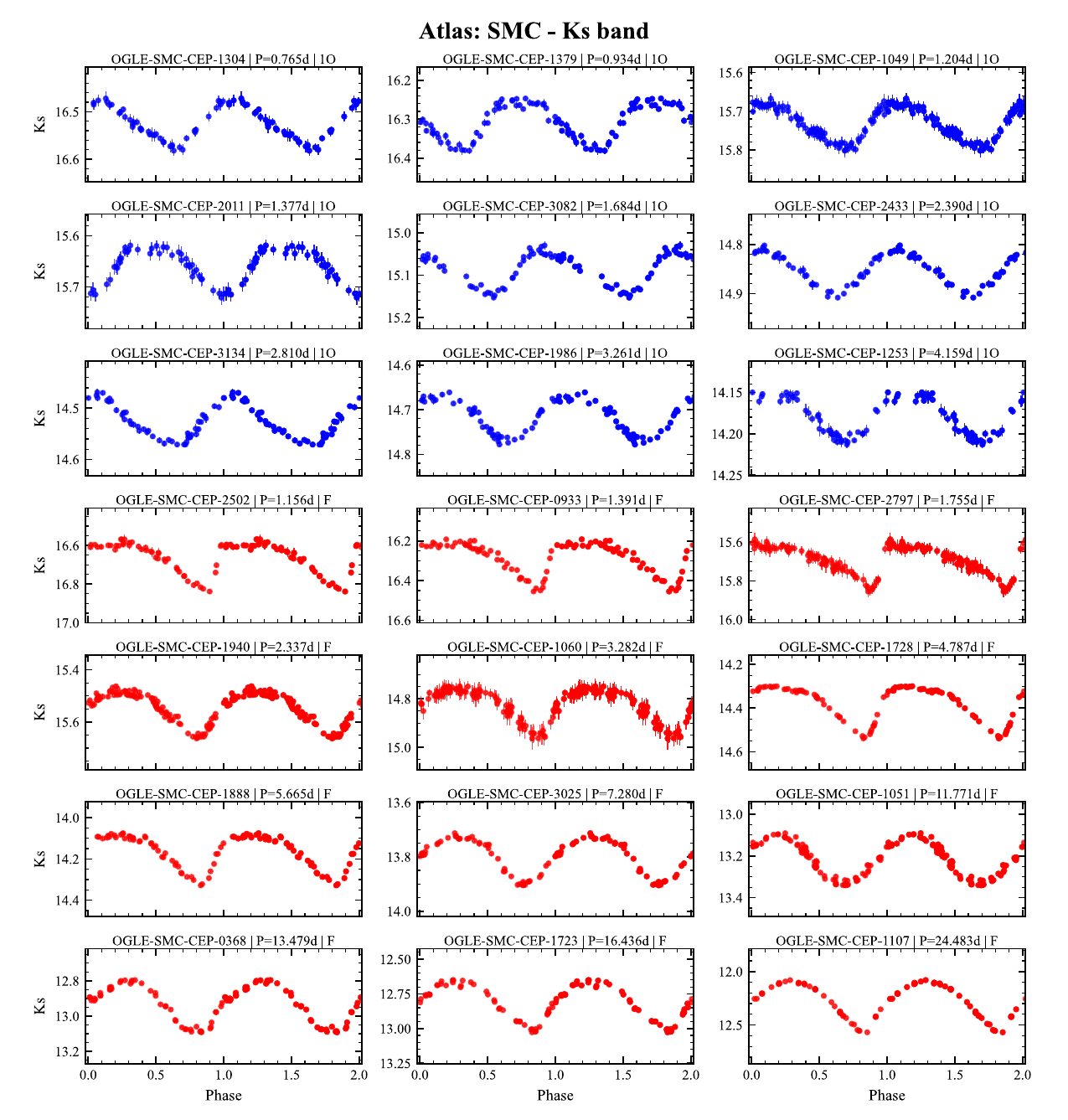}
       \caption{As in Fig.~\ref{fig:lc_LMC_J}, but for the $K_\mathrm{s}$ band in the SMC.}
          \label{fig:lc_SMC_Ks}
    \end{figure*}

 \section{PL and PW relations with fixed slopes}
\FloatBarrier

\begin{figure*}[!]
    \centering
    \includegraphics[angle=0,width=14.cm]{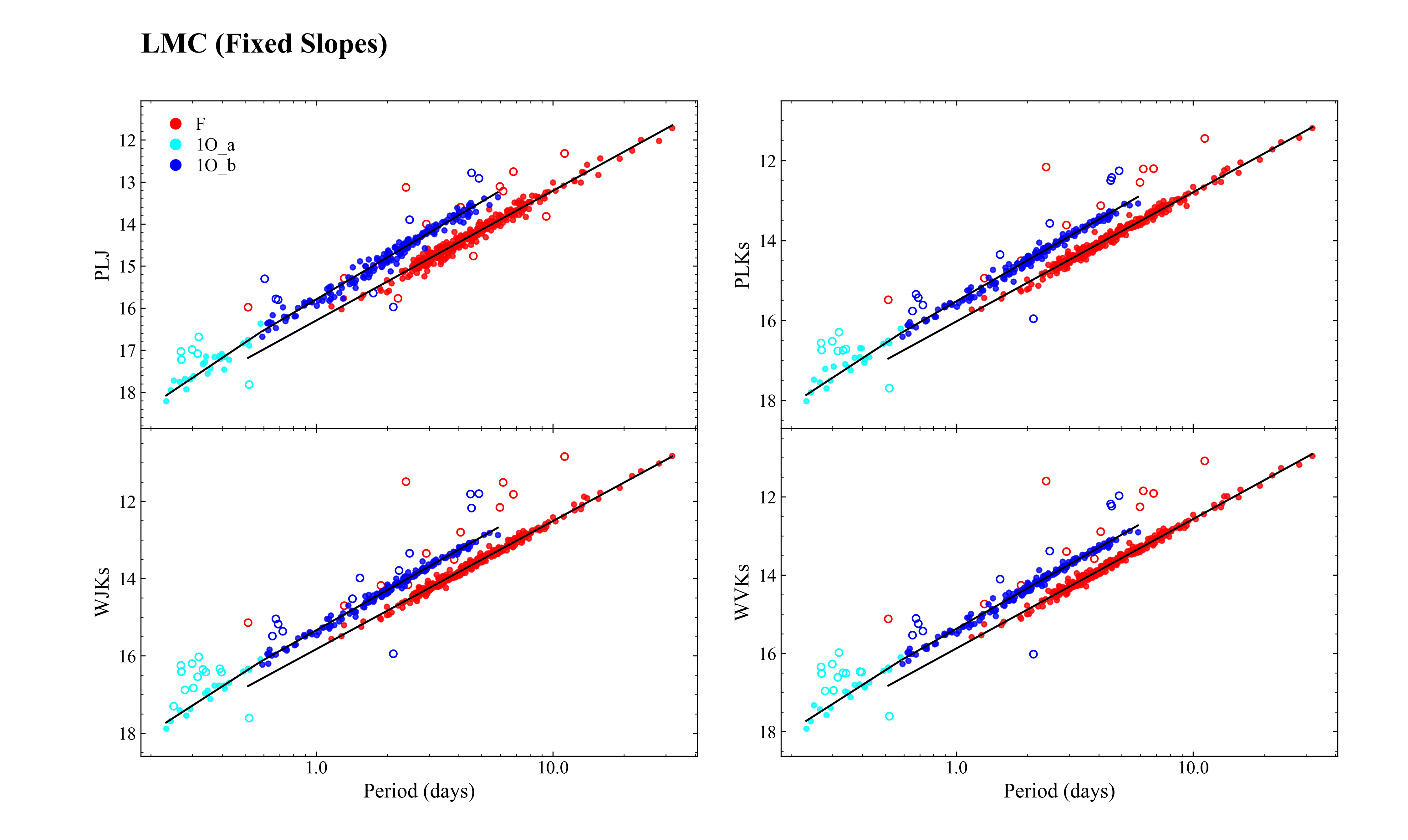}
       \caption{PL relations in $J$ and $K_\mathrm{s}$ for the LMC. Red, blue, cyan and grey filled circles represent F-mode, 1O-mode with period > 0.58d, 1O-mode with period $\leq$ 0.58d, 2O-mode, respectively. The open circles show the outlying DCEPs which are rejected by $\sigma$-clipping during the fitting procedure.}
          \label{fig:PL_LMC_fixed}
    \end{figure*}

\begin{figure*}[!]
    \centering
    \includegraphics[angle=0,width=14.cm]{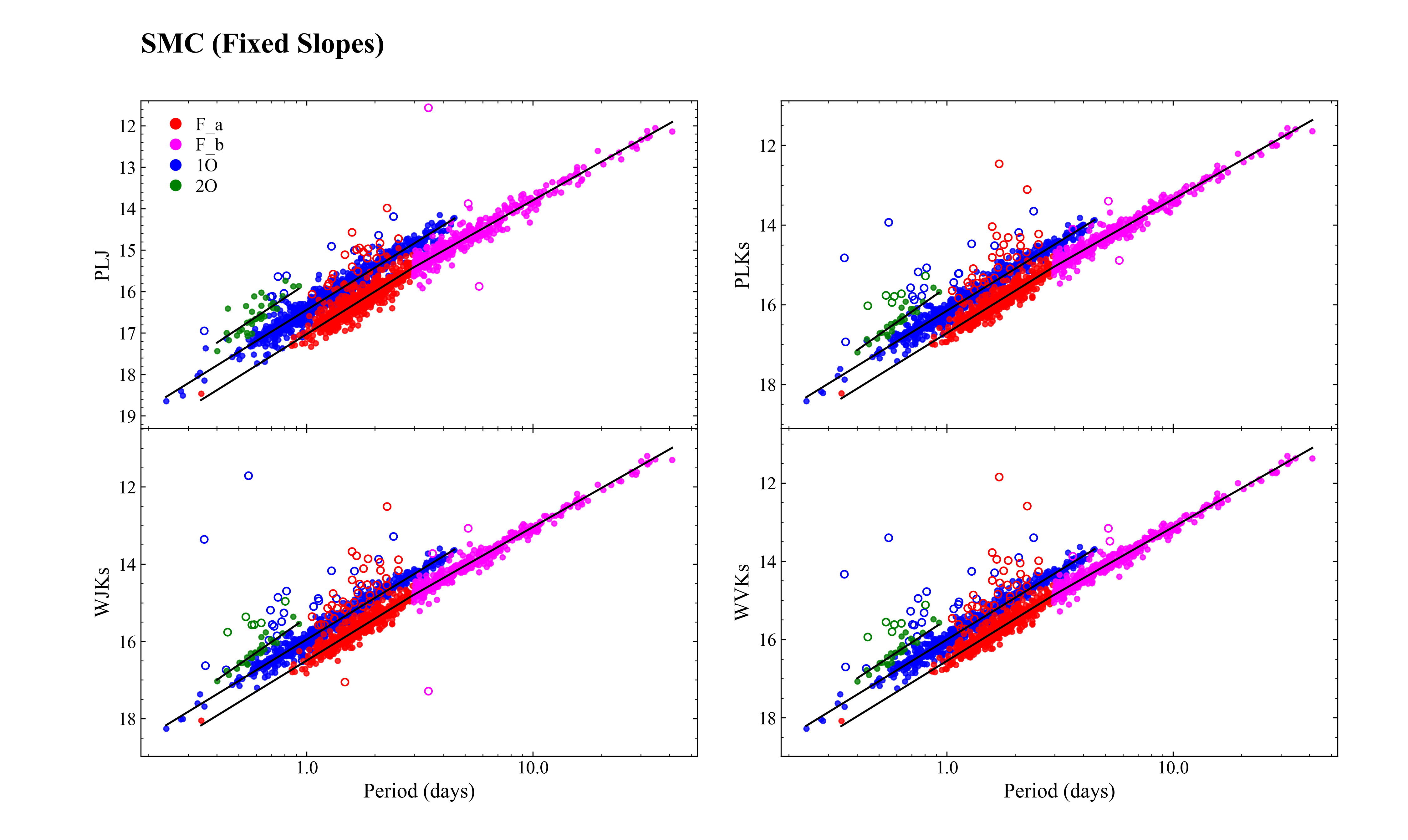}
       \caption{As in Fig.~\ref{fig:PL_LMC} but for the PL relations in the SMC.}
          \label{fig:PL_SMC_fixed}
    \end{figure*}

Figures~\ref{fig:PL_LMC_fixed} and ~\ref{fig:PL_SMC_fixed} show the PL and PW relations 
obtained with fixed slopes in the LMC and SMC, respectively. 

\section{Theory--observations comparison in the Fourier planes}
\FloatBarrier

    \begin{figure*}[!]
    \centering
    \includegraphics[angle=0,width=14.cm]{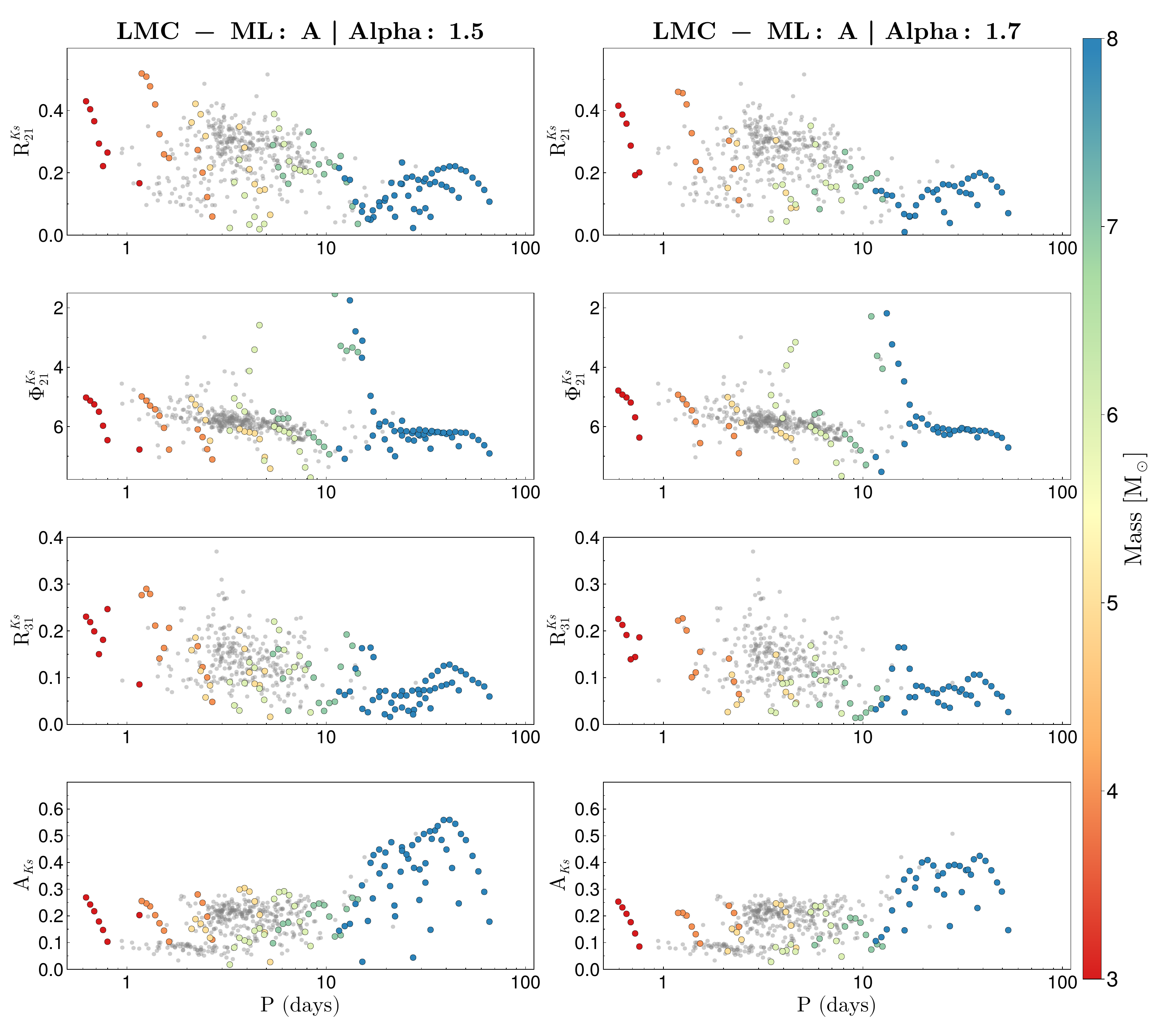}
       \caption{As in Fig.~\ref{fig:fou_lmc} but for the ML\,A models.}
          \label{fig:fou_lmc_a}
    \end{figure*}

    \begin{figure*}[!]
    \centering
    \includegraphics[angle=0,width=14.cm]{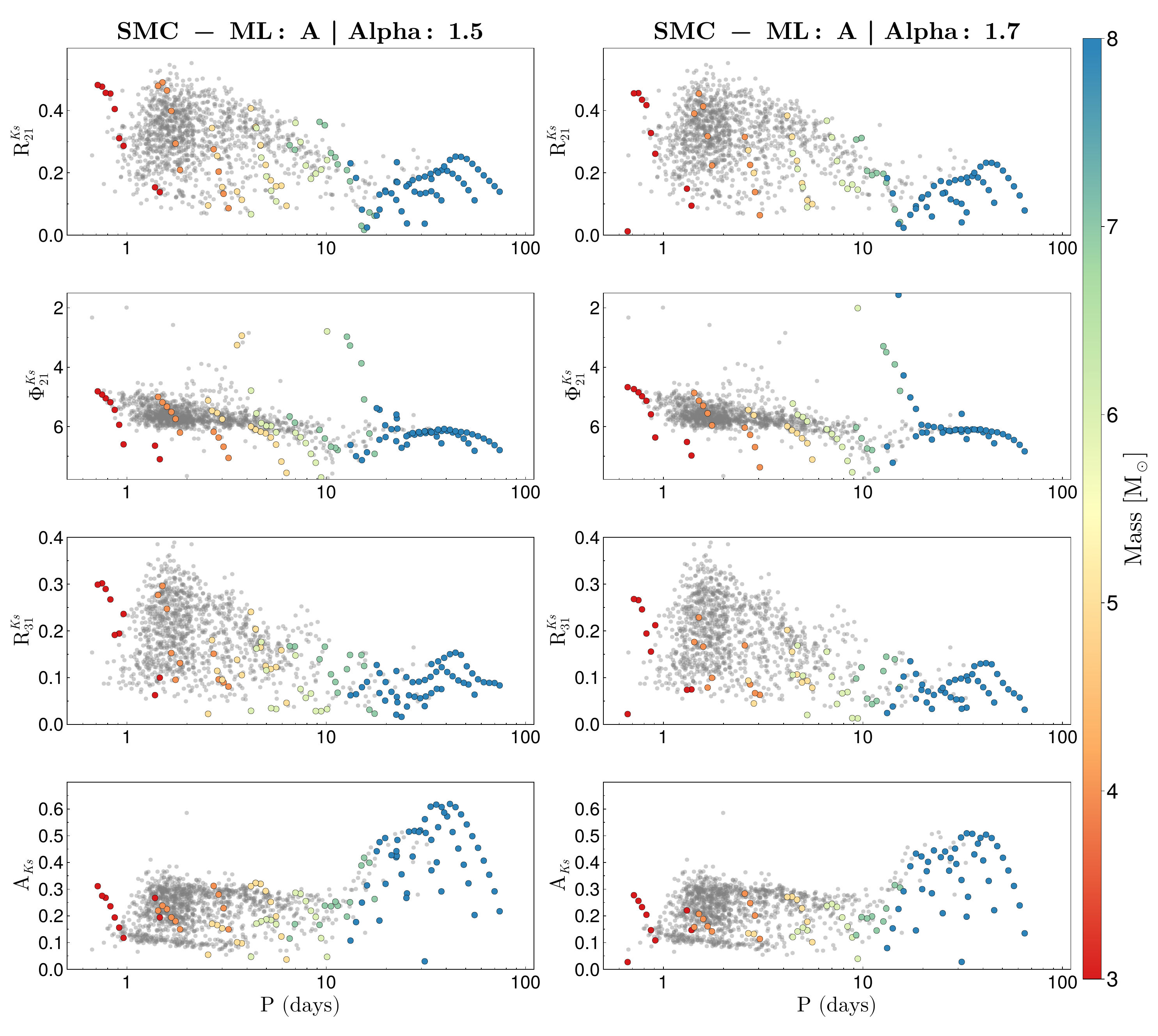}
       \caption{As in Fig.~\ref{fig:fou_smc} but for the ML\,A models.}
          \label{fig:fou_smc_a}
    \end{figure*}

\section{Error calculation on the mean age of Cepheid forming episodes}
\label{app:error}

To rigorously determine the uncertainty on the mean age of the star formation episodes, we eschew the formal covariance matrix of the least-squares fit, which typically underestimates the true error by assuming a perfectly parameterised model and ignoring binning systematics. Instead, we estimate the total uncertainty on the mean ($\sigma_{\text{tot}}$) by adding two data-driven error components in quadrature:
 $\sigma_{\text{tot}} = \sqrt{\sigma_{\text{boot}}^2 + \sigma_{\text{shift}}^2}$. 
The first component, $\sigma_{\text{boot}}$, represents the robust statistical uncertainty derived via a bootstrap resampling technique. We generated $N = 1000$ mock data sets by drawing ages from the empirical distribution with replacement, fitted each realisation with the corresponding Gaussian models, and adopted the standard deviation of the resulting means as $\sigma_{\text{boot}}$. The second component, $\sigma_{\text{shift}}$, accounts for the arbitrary choices in the discretisation of the data. We refitted the distributions across a grid of varying histogram parameters, spanning different total bin numbers and applying slight offsets to the bin edges. The standard deviation of the resulting fitted means provides $\sigma_{\text{shift}}$, quantifying the sensitivity of our peak age determination to the chosen histogram architecture.

 \end{appendix}

 \end{document}